\documentstyle[11pt,aaspp4,psfig]{article}
\begin{document} 

\title{\Large THE EMERGENCE OF THE MODERN UNIVERSE: \\ 
TRACING THE COSMIC WEB \\
\normalsize (23 June 1999) }
 

\vspace{1cm}

\begin{center}
{\bf White Paper of the UV-Optical Working Group (UVOWG) }
\end{center}
 
\author{J. Michael Shull$^1$, Blair D. Savage$^2$, Jon A. Morse$^1$ \\
Susan G. Neff$^3$, John T. Clarke$^4$, Tim Heckman$^5$, 
      Anne L. Kinney$^6$ \\
Edward B. Jenkins$^7$, Andrea K. Dupree$^8$, Stefi A. Baum$^6$, 
      and Hashima Hasan$^9$ }

\vspace{1cm}

\noindent
$^1$ University of Colorado, Dept.\ of Astrophysical \& Planetary Sciences  \\
$^2$ University of Wisconsin, Astronomy Department \\
$^3$ NASA--Goddard Space Flight Center, Laboratory for Astrophysics \& Solar Physics \\
$^4$ University of Michigan, Dept.\ of Atmospheric \& Oceanic Sciences  \\
$^5$ Johns Hopkins University, Dept.\ of Physics \& Astrophysics \\
$^6$ Space Telescope Science Institute \\
$^7$ Princeton University, Dept.\ of Astrophysical Sciences \\
$^8$ Smithsonian Astrophysical Observatory \\
$^9$ NASA Headquarters and Space Telescope Science Institute \\

\vspace{2cm}

\noindent
Because the figures for this paper are quite large, we have not included them
in this file.  The entire document, including all figures, 
may be obtained as a postscript file (or in HTML) from 
http://casa.colorado.edu/$\sim$uvconf/UVOWG.html

\newpage

\begin{table}[h]
\begin{center}
Table of Contents \\
\ \\
  \begin{tabular}{ccl}
  \hline \hline
{\it Page}   &  {\it Section}   &  {\it Topic}   \\  
\hline
3        &           &  Abstract \\
4        &           &  Preface: Background on Study and Process  \\
5        & {\bf 1}   &  {\bf Introduction}     \\
7        & {\bf 2}   &  {\bf Emergence of the Modern Universe} \\
8        &  2.1      &  Dark Matter and Baryons  \\
13       &  2.2      &  Origin and Chemical Evolution of the Elements \\
19       &  2.3      &  The Major Construction Phase of Galaxies and Quasars \\
31       &  2.4      &  Other Scientific Programs \\
38       & {\bf 3}   & {\bf UV-Optical Mission Concepts for the Post-HST Era }\\
39       &  3.1      &  Class-I Mission Concepts (4-meter telescopes) \\
46       &  3.2      &  Class II Mission Concept (8-meter telescope) \\
48       &  3.3      &  Pathfinder Mission \\
48       &  3.4      &  Additional Missions \\
50       & {\bf 4}   & {\bf Technology Roadmap}  \\
50       &  4.1      &  Overview  \\
51       &  4.2      &  Detectors \\
55       &  4.3      &  Large Light-weight Precision Mirrors \\
59       &  4.4      &  UV-Optical Components and Coatings \\
63       &  4.5      &  Summary \\
64       &           &  {\bf Appendix 1 -- Report to GST Second-Decade Committee} \\
\hline
\end{tabular}
\end{center}
\end{table}

\newpage

\begin{center}
{\Large {\bf UV-Optical Working Group (UVOWG) White Paper } }
\end{center}

\begin{abstract}

This is the report of Ultraviolet-Optical Working Group (UVOWG) 
commissioned by NASA to study the scientific rationale for new
missions in ultraviolet/optical space astronomy approximately
ten years from now, when the {\it Hubble Space Telescope} (HST) is
de-orbited.  Building on scientific talks at the August 1998
Boulder meeting, the UVOWG discussed the outstanding unsolved
scientific problems that can be answered by high-throughput UV 
spectroscopy and wide-field UV/O imaging.  Following the
exciting next decade of studies by HST (STIS, ACS, COS, WFC-3) 
and new surveys by MAP (microwave background at $1^{\circ}$ scale),
the Sloan Survey, and GALEX (discovery of over $10^6$ new QSOs 
for background targets), the stage is set for cosmological 
explorations of galaxy assembly
and the evolution of the intergalactic medium (IGM). 
Realizing that a major new UV/O mission would 
produce forefront science in all areas of modern astronomy, the
UVOWG focused on a scientific theme,
{\it ``The Emergence of the Modern Universe''}, 
that unifies many of the unsolved problems in UV/O astronomy.
We define this era as the period from redshifts $z \approx 3 \rightarrow 0$, 
occupying over 80\% of cosmic time and beginning after the first 
galaxies, quasars, and stars emerged into their present form. 
The exciting science to be addressed in the post-HST
era includes studies of dark matter and baryons, 
the origin and evolution of the elements, and the major construction 
phase of galaxies and quasars. Key unanswered questions include: 
Where is the rest of the unseen universe?
What is the interplay of the dark and luminous universe?
How did the IGM collapse to form the galaxies and clusters?
When were galaxies, clusters, and stellar populations
assembled into their current form?
What is the history of star formation and chemical evolution? 
Are massive black holes a natural part of most galaxies?
 
A large-aperture UV/O telescope in space (ST-2010) will provide a major
facility in the $21^{\rm st}$ century for solving these scientific problems.   
The UVOWG recommends that the first mission be a 
4-m aperture, SIRTF-class mission that focuses on UV spectroscopy and 
wide-field imaging.  In the coming decade, NASA should investigate
the feasibility of an 8-m telescope, by $\sim2010$,
with deployable optics similar to NGST.  The UVOWG recognizes that,
like SIRTF and NGST, no high-throughput
UV/Optical mission will be possible without significant NASA investments 
in technology, including UV detectors, gratings, mirrors, and imagers.
To achieve our science goals, the ST-2010 spectrograph will need to 
deliver over 100-fold increase in throughput and multiplex efficiency.
Likewise, the ST-2010 imagers should achieve similar gains in field
of view and efficiency.

\end{abstract}

\newpage

\section*{Preface: Background on the Study and Process}
 
This document represents the ``White Paper on UV-Optical Space
Astronomy'' commissioned by the Office of Space Science at NASA.
The scientific ideas and mission concepts grew out of
a NASA-sponsored conference held in Boulder, CO between August 5--7,
1998, entitled  {\it Ultraviolet-Optical Space Astronomy beyond HST}.
On the final day of that conference, a panel discussion focussed on
mission priorities for the period approximately ten years from
now, when the {\it Hubble Space Telescope} (HST) completes its 20-year
lifetime and is de-orbited.  At this time,
the astronomical community will lose capabilities for
high-resolution imaging and spectroscopy with a stable point-spread
function, high dynamic range, and wide field of view in the UV and optical.
 
This panel formed a strong consensus for a mission whose main
focus was high-throughput UV spectroscopy and wide-field imaging.
The most credible mission concept was one whose core science
included cosmological studies of the major epochs of galaxy assembly,
the large-scale structure of the intergalactic medium (IGM), and the
origin and chemical evolution of heavy elements.  Addressing even the
initial (small) portions of these science goals would require an enormous
dedication of HST observing time (over 10,000 orbits)
to key or legacy projects, as we outline in Appendix 1.
Moreover, even the new HST instruments soon to be installed cannot break
through the required thresholds of source sensitivity or field of view.
The panel concluded that the scientific goals of the next UV/O mission
required at least an order-of-magnitude advance in UV spectroscopic
throughput, compared to HST with the Cosmic Origins Spectrograph (COS).  
In addition, imaging in the optical and UV will be important 
for studies of galaxy assembly and related areas, if one can
achieve wide-field detector formats to facilitate efficient
measurements at cluster and supercluster scales.  Wide fields
are also critical for studies of star-forming and nebular regions
in the Milky Way and nearby galaxies.

Subsequent to that meeting, NASA requested further study of
these issues, and chartered the UVOWG (``Ultraviolet - Optical
Working Group'').  The UVOWG will report to NASA and its advisory groups,
the subcommittees for {\it Origins} and {\it Structure and Evolution of
the Universe}.   The UVOWG was chaired by Mike Shull (Colorado), and
its core membership included Blair Savage (Wisconsin), Susan Neff
(NASA-Goddard Space Flight Center), Anne Kinney (Space Telescope Science
Institute), John Clarke (Michigan), Ed Jenkins (Princeton), Tim Heckman
(Johns Hopkins), Andrea Dupree (Harvard-Smithsonian Center for
Astrophysics), and {\it ex officio} members Hashima Hasan (NASA HQ and STScI),
Stefi Baum (STScI), and Jon Morse (Colorado).  Additional contributions
to our discussions and report were made by Harley Thronson (NASA HQ),
Harry Ferguson, Melissa McGrath, Marc Postman, Megan Donahue, Carol Christian,
Steve Beckwith (STScI), Ken Sembach, Julian Krolik,
Zlatan Tsvetanov (Johns Hopkins), Bruce Balick (Washington),
Todd Lauer (NOAO), Scott Trager (Carnegie Observatories),
Bruce Woodgate, Randy Kimble, Chuck Bowers, Ritva Keski-Kuha,
David Leckrone (NASA-Goddard), James Green, Erik Wilkinson,
Jeffrey Linsky, Brad Gibson, Mark Giroux, John Bally (Colorado),
Chris Martin, David Schiminovich (Caltech), Ossy Siegmund (Berkeley),
Bob Woodruff (Ball Aerospace),
and Charles Lillie (TRW).  Many others provided comments and reactions
to our work, and we thank them for their efforts.
 
\section{INTRODUCTION}

The {\it HST and Beyond} (Dressler) Committee emphasized that
HST will be unique for conducting UV studies during the
first decade of the $21^{\rm st}$ century. They 
recommended that HST be operated past 2005, with an emphasis
on ultraviolet spectroscopy and wide-field, high-resolution
optical imaging.  The second decade of HST has great promise, as 
powerful new spectrographic (STIS, COS) and imaging instruments (ACS, WFC-3)
are installed and commissioned.  Our committee considered the 
primary science that HST may do, through a system of key projects
or ``legacy'' projects (Appendix 1).  However, in thinking about the ten-year
HST horizon, the UVOWG and others in the UV-Optical community
have been forced to confront the limitations of HST for solving
new scientific issues. To solve the critical science problems posed 
in this report, it is essential to have a highly capable UV/Optical
Observatory (ST-2010) shortly after HST is de-orbited. 
 
Space astronomy has made enormous scientific advances in
the 1990s with the imagers and spectrographs aboard the
{\it Hubble Space Telescope}.  However, UV spectroscopy has
not kept pace with ground-based scientific instruments,
which are revolutionizing studies of galaxies, quasars,
star-forming regions, and cosmology.
Forefront spectroscopic studies of galaxies, stars,
quasars, and the universe cannot be done in the $21^{\rm st}$
century with a 2.4-meter telescope.
The proposed mission or missions, referred to under
the title ST-2010, promise to fulfill
the 1946 vision of Lyman Spitzer for a {\it ``large
reflecting satellite telescope, possibly 200 to 600 inches
in diameter''} whose major scientific missions would be to study
the {\it ``extent of the universe, the structure of galaxies,
the structure of globular clusters, and the nature of other planets''}
(see Spitzer 1990).  To this list, we might add, 
with 50 years of hindsight, the
nature of quasars, the formation of galaxies, and the cosmological
processes that govern the formation of large-scale structure of
the intergalactic gas and dark matter.
 
The reasons for planning large-aperture (4-8 meter)
telescopes in space have not changed since Spitzer's visionary
report. Cosmology is inherently a photon-starved endeavor,
particularly when one requires significant spectral resolution
to detect the key astrophysical diagnostics of structure,
velocity, and composition.  Emission-line regions that probe
the assembly of galaxies and quasars are distant and faint.
Viable background targets (hot stars, quasars, clusters) are faint
and rare.  Consequently, our panel highlighted the need for
high performance, to be achieved through a combination of
larger aperture, advances in optical and detector efficiency,
multiplexing, and orbits that permit high operational efficiency. 
 
A 4--8 meter telescope in space (ST-2010) will provide a major
facility in the $21^{\rm st}$ century for solving a wide range of
scientific problems in cosmology, galaxy formation,
stellar evolution, and the origin and evolution of structure
and chemical composition.  An aperture this large should be used for
forefront science in both spectroscopy and imaging.
Diagnostic spectroscopy is at the heart of astrophysical inference, while
imaging offers enormous opportunities for new discovery
and inspiration.  To be an effective instrument in the
world of $21^{\rm st}$-century astronomy, the ST-2010 spectrographs must
deliver a sizeable increase in throughput and multiplex efficiency,
over that of any of the instruments aboard HST.  
With detector arrays growing rapidly in size, the ST-2010 imagers 
should achieve similar large gains in size and efficiency.

In the ``Mission Concepts'' section of this report, the UVOWG recommends 
that NASA plan for a 4m mission, with a SIRTF-class cost envelope
($\sim$ \$400M).  An 8m-class mission should be
studied to follow, with a giant leap in technology of
mirrors and detectors.  However, one  
key recommendation of our report surpasses all others.
The UVOWG recognizes that no UV/Optical mission of either 
4m or 8m class will be possible without significant gains in technology,
including UV detectors, gratings, mirrors, and imagers.
As with SIRTF and NGST, we believe that NASA should significantly boost
its investment in UV/O technology over the next five years,
to ensure that ST-2010 can fulfill the ambitious science
goals laid out in this report.
 
ST-2010 will produce forefront science in
nearly all areas of modern astronomy, from extragalactic to Galactic
and planetary arenas, as described in later portions of this report.
To highlight the most compelling science, we focus
on a major scientific theme that unifies many of the unsolved
issues -- {\it ``The Emergence of the Modern Universe''}.  This theme
refers to the study of the era from $z \approx 3 \rightarrow 0$,
beginning just after the first galaxies and stars took  
their present form and quasar and starburst activity reached their
peak.  Included in this theme are such questions as:
\begin{itemize}
 
\item {\it Where is the rest of the unseen universe?}
 
\item {\it What is the interplay of the dark and luminous universe?}
 
\item {\it How did the IGM collapse to form the galaxies and clusters?}

\item {\it What is the role of star-formation ``feedback'' in 
   radiation and energy on galaxies?}
 
\item {\it When were galaxies, clusters, and stellar populations
     assembled into their current form?}

\item {\it Are massive black holes a natural part of most galaxies?}
 
\item {\it What is the history of star formation and galactic
chemical evolution?}
 
\item {\it How do solar systems form, and what do they look like?}  
 
\end{itemize}
These questions and projects satisfy the human desire to understand
where we came from, and how planets, stars, and galaxies formed and
evolved.  To provide more specifics on the scientific goals contained within
this general theme, we have chosen to highlight three major areas:
\begin{itemize}
 
\item {\bf Dark Matter and Baryons}
 
\item {\bf The Origin and Chemical Evolution of the Elements}
 
\item {\bf The Major Construction Phase of Galaxies and Quasars }
 
\end{itemize}

\noindent
Section 2 of this report discusses these topics in more depth,
together with other scientific programs that would be enabled by
the missions contemplated.

\section{EMERGENCE OF THE MODERN UNIVERSE }

\noindent
The UVOWG was specifically commissioned to perform the following tasks:
\begin{itemize}
 
\item Develop high-priority science goals for UV-Optical space astronomy
for the period approximately a decade from now (2010).
 
\item Summarize a plausible mission or missions that will be capable
of carrying out the high-priority science program.
 
\item Outline a technology roadmap that might plausibly lead to a
``New Start'' for construction of the recommended missions.
 
\end{itemize}

To capture the power of the ST-2010 mission most dramatically, we 
highlight the key science in the theme 
``{\it The Emergence of the Modern Universe}'' -- studies of planets, stars,
galaxies, and large-scale structure  
from redshifts $z = 3$ down to the present epoch.   This
period occupies over 80\% of cosmological time, and captures
both the origin and evolution of nearly
every major astronomical structure, from galaxies and clusters
to quasars and the IGM. During this
time (8--10 Gyr), large-scale structure developed in both
dark matter and baryons.  Much of the intergalactic gas
collapsed onto galaxies, over 90\% of the heavy elements were formed,
and the energy sources of radiation, hot gas, and dynamic
outflows acted back on the surrounding galaxies and gas clouds.
This ``feedback'' of galaxy formation is the least
understood aspect of galaxy formation and assembly.  The proposed 
ST-2010 mission will provide definitive measurements, with unprecedented 
accuracy, of the key epochs from $z = 3$ to the present, to
fill in the evolutionary gap of galaxies, the IGM, and large-scale 
structures from infancy to maturity.  This mission will also provide 
accurate low-$z$ templates required to understand the high-$z$ phenomena.

ST-2010 will connect the high-redshift universe observable by NGST 
to the low-redshift universe, which will be studied in detail by 
quantitative spectroscopy and wide-field imaging.  
If properly designed, ST-2010
will also relate near-infrared observations of the distant (NGST) 
to optical and ultraviolet observations of the local universe.
The NIR view provided by NGST is related 
to the visible and ultraviolet view of the ``local universe,''  
since the rest-frame ultraviolet light of the first galaxies is 
redshifted to 1--5 $\mu$m by cosmological expansion. 
For cosmological studies of the distant universe, $21^{\rm st}$-century 
space astronomy must take a multi-spectral view, in which NGST 
and ST-2010 complement one another, both in wavelength and across
cosmological time.  

The rest-frame ultraviolet is a particularly crucial wavelength regime, 
since it contains the dominant emission from massive hot stars and 
sensitive resonance transitions of nearly all abundant atoms and
ions.  For the era that ST-2010 will study ($z = 0 \rightarrow 3$),
most of these transitions will be observed in the spectral region 
between about 1200 and 6000 \AA.  In absorption, these lines provide 
a unique, sensitive diagnostic for first detecting and then 
investigating warm and cool 
intergalactic gas (which may well comprise the major cosmic repository 
of baryons). In emission and absorption, these transitions allow one 
to study the dynamics, chemical composition, and physical conditions 
in environments ranging from the atmospheres and winds of stars, to 
protostellar outflows, to galactic disks, halos, and the IGM.

\subsection{Dark Matter and Baryons}

The current state of affairs in cosmology is both fluid and
exciting.  Just around the corner are powerful new techniques 
(observations of Type Ia supernovae at high redshift) 
and new missions (the MAP microwave 
background explorer) that should obtain accurate measurements of
fundamental cosmological parameters, such as the Hubble expansion rate
($H_0$), the age of the universe ($t_0$), and the contributions 
of baryons, dark matter, and vacuum energy to spacetime curvature.  
The cosmological parameter $\Omega$ measures the ratio of the
density of gravitating matter (or energy) to the critical value
needed to halt the Hubble expansion.  Various types of matter  
are labeled by subscripts:  $\Omega_b$ measures baryonic matter,
$\Omega_0$ measures matter of all types, and $\Omega_{\Lambda}$
measures an exotic new form of ``vacuum energy'' associated with
a repulsive cosmological constant.   

Current measurements of mass and space curvature suggest that
we live in a universe dominated by dark matter, and
possibly permeated by a mysterious vacuum energy.  A working hypothesis
has recently developed in which the various contributions to the
closure density have the approximate values: total matter 
$\Omega_0 \approx 0.3$ and vacuum energy (cosmological constant) 
$\Omega_{\Lambda} \approx 0.7$. Nucleosynthetic modeling of 
D/H (Burles et al.\ 1999) suggests that the baryons 
contribute only $\Omega_b \approx 0.045$
(for $H_0 \approx 65$ km~s$^{-1}$~Mpc$^{-1}$), which requires that
a substantial amount of the mass density is dark and 
probably non-baryonic.  
In ten years, these parameters may be well determined, and the major
challenge will be to identify the exact forms of the various
components of mass/energy in the universe.  An even more fundamental
physical issue is to address the question of how and why these parameters
took their current ratios.  For example, what 
sets the ratios $\Omega_0 / \Omega_b \approx 6.6$ and
$\Omega_{\Lambda} / \Omega_0 \approx 2.3$?  What is the form of the
exotic dark matter, and how is it distributed relative to 
the galaxies and the gaseous IGM?  

Ultraviolet spectroscopy (mapping the Ly$\alpha$ forest
on sub-degree scales) and wide-field imaging (weak gravitational 
lensing on cluster and supercluster scales) can address some of 
these fundamental issues. At moderate redshifts, $z < 1.5$, quasar 
absorption spectra contain evidence for the epochs of galaxy formation, 
metal production, reionization, and reheating of the baryons left over 
from the Big Bang.  Theories of primordial nucleosynthesis and 
cosmological structure formation predict a distributed IGM   
containing a substantial fraction of the hydrogen and helium synthesized
in the Big Bang. According to cosmological N-body hydrodynamic models
(Cen et al.\ 1994; Hernquist et al.\ 1996; Zhang et al.\ 1997) 
gas in the high-redshift IGM begins to collapse into the filamentary
web of dark-matter potential wells.  The first collapsed objects
(``proto-galaxies'') may form between redshifts $z = 10-20$,
and the first galaxies and QSOs are probably present by $z = 5-10$.  
As far as we can tell, the universe at redshift $z > 5$
is nearly opaque at UV and optical wavelengths,
owing to the strong absorption from hydrogen Ly$\alpha$ in the
IGM (the Gunn-Peterson trough).

The era at $z > 5$ has often been termed the ``dark ages''.
Probing these dark ages forms one of the key goals of 
NGST.  The NGST hopes to detect the first stars, first galaxies, 
and first supernovae in their redshifted light 
between 1 and 5 $\mu$m.  ST-2010 will observe the fruits of 
these seeds of galaxy formation, to see ``how it all turned out''.  
The type of object that NGST will study at $z > 5$ needs to be
characterized in the rest-frame ultraviolet, especially in the 
low-redshift ``modern universe'',
where we can obtain high-resolution UV/Optical images, UV spectra, 
and large-scale maps of the distributions of galaxies 
and intergalactic clouds. 
 
Thus, the premier challenge for UV astronomy in 2010 will be to 
follow the evolution of the universe from the ``dark ages'' down to the
``renaissance'' of star formation, supermassive black holes (quasars),
and metal production in the present epoch.  
Measuring the evolution of the Ly$\alpha$ baryon content is
vital if we are to understand the mass evolution of galaxies, the
rate at which gas in the IGM is incorporated into galaxies, and
the rise of metallicity over cosmic time.
At low redshifts ($z < 1$) new astronomical instruments on ST-2010
will be able to see these objects in remarkable clarity.
To elucidate the emergence of the modern universe in the gas, stars, 
and galaxies, NASA needs to set into motion the technological 
development that will make ST-2010 ready for a New Start late next decade.   

\vspace{0.5cm}
\noindent
{\bf Mapping the Large-Scale Structure of the IGM} \\

\begin{figure}[h]
  \caption[]{Large-scale cosmological structure, consisting of filaments 
   of galaxies surrounding voids, is seen in the CfA2 redshift survey 
   (Huchra 1999).  This ``pie-diagram'' shows the distribution 
   in recession velocity and right ascension of bright  galaxies and 
   four Ly$\alpha$ absorbers found by HST/GHRS toward Mrk~501 and Mrk~421 
   (Penton, Stocke, \& Shull 1999).  Evidently, the voids are not entirely
   empty:  two Ly$\alpha$ clouds lie in voids, with the nearest bright 
   galaxies more than 4 Mpc away. } 
\label{pie}
\end{figure}

Based on recent galaxy redshift surveys, astronomers have detected the
existence of an organized large-scale structure in the galaxy
distribution, which takes the form of large filamentary walls
and ``empty'' voids.  By 2010, these galaxy surveys will outline
the distribution of luminous matter in fine detail, but the
dark, gaseous universe (the IGM) will remain largely unexplored
at $z < 1.65$. (At $z > 1.65$, the Ly$\alpha$ line is redshifted
into the visible band, although several key metal transitions at $\lambda <
1216$~\AA\ remain in the ultraviolet.)   Theoretical models suggest that
studies of the H~I and He~II Ly$\alpha$ forest of absorbers in
QSO spectra should probe the large reservoir of
gas left from the major epoch of structure formation.  In fact, the
intergalactic Ly$\alpha$ absorbers persist down to very low
redshifts, and observations from HST show that many Ly$\alpha$
clouds exist in voids as well as in filamentary walls (see Fig.~\ref{pie}).

Studies of the He~II Ly$\alpha$ forest are particularly 
effective at probing the lowest-density regions of
the baryon distribution, while the H~I Ly$\alpha$ lines
at redshifts $z < 1.65$ (Bahcall et al.\ 1996; Stocke et al.\ 1995;
Jannuzi et al.\ 1998; Weymann et al.\ 1998)
may be used to follow the hydrogen structures down to the present epoch.  
In combination, these two diagnostics allow astronomers to follow the
interplay between the formation of galaxy structures and the IGM.
They can also be used to 
study mass exchange:  the depletion of the reservoir of 
intergalactic gas into galaxies, and the flow of mass from galaxies 
back to the IGM through galactic winds and tidal stripping.

\begin{table}[h]
  \begin{center}
Table 2.1: Quasar Number Counts and the Mean Angular Distance Between QSOs \\
\ \\
  \begin{tabular}{ccc}
  \hline \hline 
  $m_B$      &   $N_{\rm QSO}$   &  $\theta_{\rm QSO}$  \\
  (magn)     &   (sqdeg$^{-1}$)  &   (arcmin)           \\
\hline
16  &  0.01  &  $300'$   \\
17  &  0.13  &  $ 83'$   \\
18  &  1.1   &  $ 29'$   \\
19  &  5.3   &  $ 13'$   \\
20  &  17    &  $7.3'$   \\ 
21  &  41    &  $4.7'$    \\
\hline
\end{tabular}
\end{center}
\end{table}
 
Because of the steepness of the quasar luminosity function,
particularly in the UV, a factor 10 better sensitivity will open up 
50--100 times more background AGN targets to probe the IGM and galaxy halos
at intermediate and low redshift.    
With these UV background sources, we can make tomographic maps of the full
``cosmic web'' (Cen \& Ostriker 1999) of the filamentary 
distributions of hot (shocked) and warm (photoionized) baryons 
left over from the epoch of large-scale 
structure formation.  This theoretical prediction is 
robust among many models, but the structures have not yet been detected.
Mapping the evolution of these gaseous structures
down to low redshift will be a prime scientific goal of ST-2010.  
\begin{figure}[ht]
   \caption[]{Large-scale filamentary structures in both hot and
    warm baryons are predicted in numerical N-body hydrodynamical
    models (cf., Cen \& Ostriker 1999). Mapping these structures in
    absorption requires ST-2010 spectra toward background quasars 
    at $m_B = 19-20$.}  
\label{cen}
\end{figure}

The Sloan Digital Sky Survey (SDSS) modeled $N_{\rm QSO}$, the expected QSO 
number counts per square degree, using data from Crampton et al.\ (1987) and 
La~Franca \& Cristiani (1997). If the sources are distributed randomly,
with mean value $N_{\rm QSO}$ (QSOs per square degree), then the mean 
angular distance between sources is $\theta_{\rm QSO} = 
(1/2)/N_{\rm QSO}^{1/2} =  (30')/ N_{\rm QSO}^{1/2}$.  
As shown in Table 2.1, the QSO counts rise rapidly at magnitudes $m_B > 18$.     
For suitable UV background targets, one should reduce $N_{\rm QSO}$ from 
these values, to allow for some QSOs being faint in the UV. 
To attain sufficient spatial coverage of the large-scale structures
in galaxies (Fig.\ \ref{pie}) and the IGM (Fig.\ \ref{cen}), 
we need to observe QSOs at magnitudes down to $m_B \approx 19-20$, where 
the the mean angular distance between QSOs on the sky is 20-30 arcmin,
allowing for the lower UV continuum owing to intergalactic absorption. 
After accounting for ultraviolet absorption from Lyman-limit 
systems,  Picard \& Jakobsen (1993) found a steep rate of 
increase, $d(\log N)/d(\log F_{\lambda})
= 2.7 \pm 0.1$ for quasars in the flux range $10^{-14}$ down to
$10^{-16}$ ergs cm$^{-2}$ s$^{-1}$ \AA$^{-1}$ (approximately
$m_B = 15$ down to $m_B = 20$).  The current limit
of HST/STIS for moderate-resolution spectroscopy is 
$m_B \approx 15$, while  HST/COS will take this limit
to $m_B \approx 17.5$.  Another order-of-magnitude improvement
is required to capitalize on the large increase in QSO populations
at magnitudes $m_B = 18 - 20$.  

In the next several years, the GALEX mission is expected 
to identify large numbers ($10^{5-6}$) of QSOs in the magnitude 
range $18 < m_B < 20$. 
The Sloan survey will provide redshifts for $\sim10^5$ of these
targets.  The task of mapping the IGM structures from 
$z = 2 \rightarrow 0$ will be a major highlight of ST-2010's program.
Its spectroscopic throughput is sufficient to undertake a  
major survey of sightlines at high spatial frequency.
The goal is to make an IGM baryonic survey on sub-degree 
angular scales, comparable to that of the 
MAP explorer and to the structure seen in galaxy surveys.  
In doing so, we will connect the high-redshift seeds of galaxies and 
clusters with the distributions of galaxies and IGM in the modern epoch,
at redshifts $z < 1$.   

\vspace{0.5cm}
\noindent
{\bf A Baryon Census of the Local Universe} \\ 

It has been estimated (Fukugita, Hogan, \& Peebles 1998)
that 50\% of all baryons predicted by Big Bang nucleosynthesis
may reside in undetected form at low redshift, perhaps in a hot
intergalactic medium (Cen \& Ostriker 1999), 
in photoionized H~I clouds (Shull 1998),
or in small groups of galaxies.  Identifying these ``missing baryons''
and other dark matter can be done by studying the Ly$\alpha$
absorbers, galactic halos, and weak gravitational lensing on
cluster and supercluster scales.  In fact, the Ly$\alpha$
surveys may be a better tracer of dark matter than galaxies,
since they can probe relatively uncollapsed material in the IGM.
Measuring the evolution of the Ly$\alpha$ baryon content is
vital if we are to understand the mass evolution of galaxies, the
rate at which gas in the IGM is incorporated into galaxies, and
the rise of metallicity over cosmic time.

\begin{figure}[h]
   \caption[]{Intergalactic Ly$\alpha$ clouds are ubiquitous, even
    at low redshift.  This HST/GHRS spectrum of the background source
    PKS~2155-304 shows multiple Ly$\alpha$ absorption systems between
    1281--1290~\AA, or line-of-sight (LOS) recession velocity
    $cz = 15,700 - 17,500$ km~s$^{-1}$.
    These absorbers appear to arise from clumps of intergalactic
    hydrogen gas, spread over a region $\sim1$ Mpc around a group
    of intervening galaxies (Shull et al.\ 1998).  The gas may be
    pristine in metal abundances, as shown by the absence of Si~III or
    C~IV at the velocities of the H~I absorption.  }
\label{2155}
\end{figure}

The low-redshift Ly$\alpha$ clouds provide powerful probes
of large-scale structure, since they are easy to detect and far
more numerous than bright galaxies.
Preliminary studies by HST/GHRS and HST/STIS toward 15 
bright QSO targets have shown that the low-redshift Ly$\alpha$
forest remains ubiquitous, even down to $z = 0$ (Fig.~\ref{2155}).
Moderate resolution spectroscopic surveys (Penton et al.\ 1999)
identify one low-$z$ 
Ly$\alpha$ absorber for every 1500 km~s$^{-1}$ of redshift down to column 
density N(H~I) $= 10^{12.6}$ cm$^{-2}$.  As with the Ly$\alpha$ absorbers
at high redshift, these clouds appear to have large cross sections,
of order 100 kpc in size.  However, the inferred space density of these
Ly$\alpha$ absorbers is still remarkably large, 
$\sim 0.4$~Mpc$^{-3}$, which is comparable to the density of dwarf galaxies 
and some 20 times larger than that of typical bright ($L^*$)
galaxies.  A rough estimate (Shull 1998) of the clouds' contribution to the
local mass density of the universe gives $\Omega_{\rm cl} \approx
(0.008 \pm 0.004) h_{75}^{-2}$, or some 20\% of the baryon density
consistent with D/H nucleosynthesis.

Clearly, a significant number of baryons are left in the IGM,
even at low redshift, but a more detailed baryon census is needed.
The goal of this survey would be to measure the frequency
of absorbers, per unit redshift, and the distribution in H~I
column density for a large ensemble of sightlines.  To convert
the distribution, $f(N_{\rm HI}, z)$, into a space mass density
requires two additional pieces of information:  first, the range of
sizes and topology of the clouds (their cross section), and second,
a realistic ionization correction for $n({\rm H}^+)/n({\rm H}^o$).
Crude values for these parameters went into the preliminary
estimates of $\Omega_b(z=0)$, but this is no substitute for
a large-scale tomographic survey, using neighboring sightlines
to constrain the cloud sizes and shapes.  To derive the 
evolution of $\Omega_b(z)$ from $z \approx 2$ down to the present epoch
will require major surveys of many QSOs, down to $m_V \approx 20$,
a task that exceeds the capability of HST/COS by a factor of 10.  
In addition, because of the low detector efficiency in the 
near-UV on STIS and COS, adequate surveys for 
Ly$\alpha$ absorbers at $z = 1.4 - 1.8$ are difficult.  This is
precisely the epoch when the evolution in the absorbers appears
to change drastically.   Finally, a vastly increased sample of 
Ly$\alpha$ absorbers will produce a statistically significant measure 
of the evolution of the metagalactic ionizing background radiation,
from the diminution of Ly$\alpha$ absorber frequency near quasars -- 
the so-called ``proximity effect''.  This effect is used to measure 
the metagalactic ionizing background, which is a key ingredient in the 
ionization corrections needed, both for hydrogen and 
 for accurate measurements of metallicity.

\vspace{0.5cm}
\noindent
{\bf Reionization of Hydrogen and Helium} \\ 

One workable definition of the ``end of the dark ages'' and the
emergence of the modern universe is the time when the universe becomes
largely transparent to ionizing photons (1 -- 10 Rydbergs).
The universe and IGM becomes transparent after 
the epochs when hydrogen and helium are re-ionized by the ionizing 
radiation from the first stars and first quasars.  This occurs 
at $z > 5$ for H~I, but helium is probably 
not reionized until $z \approx 3$, as inferred from 
the strong He~II Ly$\alpha$ absorption troughs
at $(304~{\rm \AA})(1+z)$ in the UV spectra of high-$z$ QSOs 
(see Fig.~\ref{heap}).

 \begin{figure}[h]
  \caption[]{Absorption in the high-$z$ He~II and H~I Ly$\alpha$ forest
    occurs in a myriad of discrete absorbers, probably arising
    from density fluctuations in the IGM.   
    The HST/STIS spectrum (solid line at bottom) of the quasar 
    Q0302-003 shows He~II absorption shortward of 
    1300~\AA\ (Heap et al.\ 1999) with superposed higher
    resolution spectrum of H~I Ly$\alpha$
    from Keck/HIRES.  The H~I Ly$\alpha$ data were  
    normalized and multiplied by 0.25 in wavelength to match the 
    He~II wavelength scale. } 
  \label{heap}
  \end{figure}

The $z = 3$ epoch may have other significance for the build-up
and transport of heavy elements throughout the IGM.  The metallicity
of intergalactic space can be measured from the strong UV resonance
lines such as C~IV $\lambda1549$, Si~IV $\lambda1400$,
C~III $\lambda977$, Si~III $\lambda1206$, and O~VI $\lambda1035$.   
These UV resonance lines are the most sensitive abundance indicators
available in astrophysics, and they are widely used as abundance indicators
in stars, in the low-redshift interstellar medium, and in the 
high-redshift IGM.
Current evidence at high redshift suggests that C~IV/Si~IV abundance 
ratios shift at $z \approx 3$, possibly due to a spectral renaissance 
stimulated by the breakthrough and overlap of the cosmological He~II
ionization fronts from QSOs and starburst sources (Songaila
\& Cowie 1996; Giroux \& Shull 1997).     
The ionizing UV radiation field from high-redshift QSOs and 
starburst galaxies is strongly filtered and processed by the
IGM (Haardt \& Madau 1996; Fardal, Giroux, \& Shull 1998).
Because helium is more difficult to ionize than hydrogen,
singly-ionized helium builds up a significant trace population,
He~II/H~I $\approx 100$, which then blocks all radiation at
energies above the He~II ionization edge at 54.4 eV.  
The 304~\AA\ (Ly$\alpha$) lines of He~II can be used to probe
low-density regions of the IGM, particularly the void-like
gaseous structures in the baryon distribution that develop
in concert with the large-scale structures in dark matter and
the reionization. 

A challenging scientific project for ST-2010 will be to probe several 
hundred high-$z$ QSOs, most at $m_B = 18-20$, that will be found by 
GALEX, SDSS, and other surveys, so that we can observe
the He~II reionization epoch in detail.  These measurements will 
also constrain the spectral nature of the ionizing sources
(QSO or starbursts) and provide the critical information for
the ionization corrections needed to obtain accurate
metal abundances and to monitor the chemical evolution rates with redshift.   
Thus, the reionization project should be done simultaneously
with the study of chemical evolution described next.

\subsection{Origin and Chemical Evolution of the Elements}

To understand the chemical evolution of the universe, we need to 
determine the elemental composition of the gaseous matter
and its relation to cosmic epoch and physical environment.   
Abundances of various species are expected to vary with gas density, 
star-formation rate, and proximity to galaxy structures. 
In the very early evolution, H, He, and trace amounts of 
other light elements such as D and Li were created in the expanding 
Big Bang fireball.
At later times, stars converted the gaseous products of the Big Bang  
into heavier elements and returned the processed elements
back to the interstellar medium via stellar winds and supernova
explosions.  Subsequently, the metal-enriched interstellar  
gas was transported to the IGM by galactic outflows, gravitational 
interactions, and mergers of galaxies.  Numerical models of IGM 
enrichment (Gnedin \& Ostriker 1997)
predict a strong dependence of metallicity on density;
the highest density regions are expected to reach 0.1 solar metallicity
at $z \approx 3$, while the matter in the voids remains nearly
pristine.  Subsequent stellar processing created new elements 
and slowly modified the nucleosynthetic imprint of the Big Bang.  
These processes of stellar element production and 
destruction have continued to the current epoch.  A study of elemental
abundances as a function of lookback time and environment
provides a detailed, quantitative assessment of the history of 
element production and destruction.

\newpage
\noindent
{\bf The Primordial Value of D/H} \\

One of the critical parameters for studies of the early universe is
the primordial D/H ratio.  By measuring D/H in a variety of environments
and following its evolution with metallicity, one can extrapolate back to
the primordial D/H, which provides an accurate measurement of the baryonic
contribution, $\Omega_b$, to the closure density (Fig.~\ref{burles}).  
By measuring D/H in a
wide variety of environments, one can understand the rate at which deuterium
is destroyed as matter is cycled through stars (``astration").
It is now recognized that, beyond the local 100 pc,
precise measures of D/H are extremely
difficult to obtain for any astrophysical site. Therefore, the ultimate goal
of determining the primordial value of D/H may become a long-standing
problem in observational astronomy, equivalent to the current quests
for obtaining accurate values of the Hubble Constant, $H_0$, and the
closure and deceleration parameters, $\Omega_0$ and $q_0$, respectively.   

\begin{figure}[h]
   \caption[]{Summary of Big Bang Nucleosynthesis predictions for
   light elements (Burles et al.\ 1999), compared to observed
   abundances of D, Li, and He.  The vertical band indicates the baryon
   density, measured in terms of $\Omega_b h^2$, inferred from recent
   D/H observations (best-fit is $\Omega_b h^2 = 0.019 \pm 0.0024$),
   where $h$ is the Hubble constant in units of 100 km~s$^{-1}$
   Mpc$^{-1}$ (recent estimates give $h = 0.6-0.8$). 
   The parameter $\eta$ is the baryon-to-photon ratio. }
\label{burles}
\end{figure}

Measuring D/H from the ground in high-redshift QSO absorption line 
systems has been difficult because of the
confusion produced by the Ly$\alpha$ forest of absorbers.  Measuring D/H
from space has been equally difficult because of the faintness of the
possible background sources and the low sensitivities of current UV
spectrographs in space. While continued progress can be expected from STIS,
FUSE, COS, and spectrographs on ground-based telescopes, a precise 
measurement
of the primordial value of D/H will require fully understanding the
formation and destruction processes affecting D and then following D/H
versus X/H as X/H $\rightarrow 0$. Here, X represents a nucleosynthetic
product of the stellar evolutionary pollution of the gas. For example, 
measures of D/H versus C/H, N/H, or  O/H would be extremely important.

Galactic studies of D with FUSE should provide new insights about the
formation and destruction processes affecting D.  Preliminary
evidence is now appearing (Jenkins et al.\ 1999) that the value of D/H
may vary by a factor of 2 in a few places in the solar region of the Milky
Way.  Such variations are difficult to understand, and they
suggest that our current knowledge of D production and
destruction in galaxies is limited.   We know even less about D/H in
matter with low metallicity, because QSO absorption-line systems
suitable for obtaining precise measures of D/H are rare.  Unless
astronomers are exceptionally lucky over the next 10 years, it is very
likely that an accurate measurement of the primordial value of D/H will
require a UV spectroscopic capability, beyond the HST era, 
that can produce high S/N spectra of faint QSOs with a resolution exceeding 
30,000.  To have a reasonable chance of finding a suitable QSO absorption
system to measure D/H requires observing many QSOs at $m_B = 19$;
the success rate is only about 2\% at $m_B \approx 16$.   

\vspace*{0.5cm} 
\noindent
{\bf Star Formation History} \\

The deep imaging by the {\it Hubble Space Telescope} has stimulated
interest in a search for the history of massive star formation from
redshifts $z = 5$ to the present.  The ``Hubble Deep Fields'' taken
by WFPC-2, STIS, and NICMOS have shed new light on the
evolution of the stellar birthrate, initially highlighting
$1 \leq z \leq 2$ as the interval when most of the optical starlight
was produced.  Subsequent ground-based studies of ``U-band dropout''
galaxies by {\it Keck} and of reprocessed sub-mm emission by
the {\it SCUBA} imager have challenged this interpretation.
As shown in Fig.~\ref{SFR}, the star formation rate may be rather
flat from redshifts $z = 2$ back to $z = 5$.

\begin{figure}[t]
 \caption[]{Comoving density of star formation versus redshift (Madau 1999).
{\it Left:} Unreddened values, inferred from UV continuum luminosity
densities. Dotted line shows the fiducial rate needed to generate
the extragalactic background light. {\it Right:} Dust-corrected
values, including measurements from the Hubble Deep Field (filled
pentagons) and from the $z = 4$ survey (open squares).
The H$\alpha$ determinations (filled triangles)
and {\it SCUBA} sub-mm lower limit (empty pentagon) are
added for comparison.}
\label{SFR}
\end{figure}

Although the star-formation history at $z > 5$ will be measured
by NGST, most of the stars and metals are formed more recently,
at wavelength bands accessible by HST and ST-2010.
The starlight probed by these deep surveys is generated primarily
by massive stars, although dust processing and radiative transfer
play an important role in calibrating the total star formation rates.
The rapid decrease in star formation and AGN activity at $z < 2$
parallels the development and assembly of modern galaxies and clusters,
when over 90\% of the heavy elements are produced and distributed.
ST-2010 imaging and spectroscopy can follow this star formation rate
and compare it to the rate of chemical evolution and galaxy assembly
through accurate abundances and kinematic mass determinations.
At low redshift, ST-2010 can perform detailed studies of the ``local
counterparts'' of the high-$z$ star-forming regions -- the massive stars,
OB associations, and super star clusters in the Milky Way and Local Group.

\vspace*{0.5cm} 
\noindent
{\bf Chemical Evolution and Metal Production in the Universe} \\ 
 
QSO absorption-line surveys have shown that the Ly$\alpha$ 
systems seen at high redshifts contain the bulk of the gas mass in
the universe (Madau \& Shull 1996; Weinberg et al.\ 1997).   
Moreover, this mass is comparable to the luminous (stellar)
mass in the universe at the present epoch. In the earliest interpretation
of the damped systems, it was suggested that they represent the progenitors
of present-day galaxy disks (Wolfe et al.\ 1986). While this interpretation
might still be approximately (but not universally) true, recent studies of
low-redshift examples (Rao \& Turnshek 1998) show that the galaxies 
responsible for the damped systems also include dwarf or 
low surface brightness galaxies.  
Spectroscopic studies of the damped Ly$\alpha$ systems should allow
astronomers to measure directly element abundances and kinematics in
gaseous systems that eventually evolve into galaxies. 
The combined method of QSO absorption-line studies with follow-up imaging and
galaxy identification offers a unique opportunity to study the extended 
gaseous and stellar components of non-local galaxies at the same time.
In principle, studies of a significant number of damped systems could be used
to measure accurately the build up of the elements and trace the changing
physical and kinematical properties of galaxies or their progenitors
over the redshift interval $0 < z < 4.5$.

At present, important abundance studies are
being pursued with major time allotments on large ground-based telescopes
for a wide range of elements including: O, N, Mg, Al, Si, S, Cr, Mn, Fe,
Ni, and Zn. These studies have been undertaken for damped Ly$\alpha$
systems with redshifts $1.65 < z < 4.5$ (Pettini et al.\ 1994, 1997; Lu et
al.\ 1996; Prochaska \& Wolfe 1999). The redshift range $1.65 < z < 4.5$
corresponds to cosmic lookback times from approximately 77\% to 90\% of the
age of the universe (assuming $q_0  = 0.5$).  This high-$z$ epoch is certainly 
an important time period in the history of the universe. However,
to truly understand the implications of the ground-based measurements of 
the higher-redshift systems, it will be necessary to combine them with 
analogous UV measurements over the redshift interval $0 < z < 1.65$, to
follow the Ly$\alpha$ systems over the last 77\% of lookback time. 

Unfortunately, damped Ly$\alpha$ systems at low redshift are rare,
and the background quasars are often faint. For example, the HST QSO
Absorption Line Key Project discovered only one damped Ly$\alpha$ system
in 83 sightlines (Jannuzi et al.\ 1998).  
However, a special search technique employed by Rao
\& Turnshek (1998, 1999) has substantially expanded the sample of
low-redshift damped systems. There are approximately 20 known
low-redshift classical damped Ly$\alpha$ systems (with neutral hydrogen
column densities of $2 \times 10^{20}$ cm$^{-2}$ or larger), 
including the systems found by Rao \& Turnshek (1999).  For all these 
systems, the background QSOs have relatively low-to-moderate far-UV fluxes 
and represent extremely difficult targets for high-resolution studies with 
either HST/STIS or COS.

 A first major survey of element abundances and kinematics in
low-redshift damped Ly$\alpha$ systems would realistically require a
spectroscopic resolution of $R = 30,000$ at $S/N > 30$ for approximately 100
such systems, uniformly covering the redshift interval $0 < z < 1.65$. By the
time such a program could be undertaken, the amount of ground-based
information on higher redshift damped systems will be enormous.
Therefore, the data obtained on the lower redshift systems will be even more
crucial to developing a self-consistent interpretation of the chemical and
kinematic evolution of the bulk of the neutral gas in the universe.  
To understand fully the implications of the measured abundances and
kinematics of the damped systems, it will be necessary to obtain
information on the corresponding absorbing galaxies through their luminous
emissions. Therefore, high angular resolution UV, optical, and IR imaging
studies of the fields of each QSO will be an important element in any such
program.

A significant complication in previous studies of element abundances in QSO 
absorption-line systems is introduced by the possible incorporation of 
various heavy elements into dust. The most important elements for future
studies by ST-2010 are those elements not usually found in dust,
but which have different nucleosynthetic origins.
Particularly important undepleted elements include Zn, S, P, N, and Ar.  Of
these, S, P, N, and Ar have rarely been studied from the ground since their
resonance lines usually lie in the Ly$\alpha$ forest and are often
confused by intervening absorption due to hydrogen Ly$\alpha$
absorption at other redshifts.  However, this spectroscopic confusion is
greatly reduced at lower redshifts, because of the rapid decrease in the
number of Ly$\alpha$ forest lines.

\vspace{0.5cm}
\noindent
{\bf Nucleosynthesis in Supernovae and Young SNRs}  \\ 

We have discussed the need to invest significant observing time over
the next decade in UV studies of QSO absorption-line systems,
probing the physical conditions and chemical evolution of the IGM
and the halos of galaxies over a large redshift range.
This database of absorption-line systems will
be used to determine accurate column densities, 
abundances, and kinematics of intergalactic matter at epochs
when the first galaxies were formed and the first heavy
elements were synthesized.  We will thus measure the production rates and
dissemination of heavy elements from massive stars in the
early universe via statistical studies of the
integrated light from gas and stars in galaxies and the IGM.
A complementary, though crucial tactic is to
study the nucleosynthesis yields directly in nearby supernovae and young
supernova remnants (SNRs).  Elemental abundance determinations in
supernova ejecta are essential for testing theories of nucleosynthesis 
occurring in massive stars, and, ultimately for models for the chemical 
enrichment of the ISM in galaxies.  Likewise, Wolf-Rayet stars and
asymptotic giant branch stars are important contributors to this
enrichment, and their mass-loss rates and wind abundances need to
be understood much better. 

By studying supernova remnants throughout the Galaxy and
the Local Group, we can investigate the role that      
supernovae play in the structural and chemical evolution of galaxies.
Galaxies become chemically enriched when supernovae inject the by-products
of nucleosynthesis occuring in the cores of massive stars into the
interstellar medium.  In addition, shock waves produced by 
supernovae heat the ISM, determine the
velocity dispersion of interstellar clouds, and
govern the scale height of the ISM in galaxies (McKee 1990). 
Core-collapse supernovae play a major role in enriching
the ISM in young galaxies and the surrounding IGM with heavy elements.
Thus, understanding nearby core-collapse supernovae, their abundance
yields, and energy output, leads to a better understanding
of these important processes in the early universe.
Of particular interest to the issue of chemical enrichment
of young galaxies are the young, ejecta-dominated SNRs in the LMC 
and SMC (see Fig.\ \ref{N132D}), 
which reflect nucleosynthesis in a low-metallicity regime of initial 
abundances, applicable to high-redshift galaxies.

\begin{figure}[t]
\caption[]{Central regions of the oxygen-rich supernova remnant
N132D in the LMC.  Narrow-band image ($\sim 1'$ FOV) with HST/WFPC2 
([O~II] $\lambda\lambda3727$ in blue, [O~III] $\lambda5007$ in green,
and [S~II] $\lambda\lambda6716,6732$ in red) reveals intricate 
interactions between supernova ejecta (nearly pure oxygen filaments 
in blue-green) and circumstellar environment (shocked clouds are 
pinkish-white strands).  Wide-field imaging in diagnostic emission 
lines combined with UV/O spectroscopy allow us to track the evolution 
of young SNRs and measure the elemental abundances in core-collapse 
supernova ejecta.}
\label{N132D}
\end{figure}

The fundamental processes of nucleosynthesis that take place deep
inside the cores of massive stars are hidden from view 
until the stars explode as supernovae.  Young SNRs therefore allow us to
investigate material from the cores of massive stars directly,
leading to observational tests of theories for stellar evolution
and nucleosynthesis.  Our glimpse of the uncontaminated supernova
debris lasts for at most a few thousand
years before this material mixes into the ISM.
There are only eight young SNRs known which contain
fast-moving ($> 1000$ km s$^{-1}$) optical filaments of uncontaminated debris.
Cas A in our Galaxy is the prototype of this oxygen-rich class, and
is joined by two additional Galactic remnants, two SNRs in the LMC, one in 
the SMC, and two unresolved objects in the more distant galaxies M83 
and NGC~4449.  The highly elevated abundances of oxygen, neon, sulfur, and 
other heavy elements suggest these debris originated within
the helium-burnt layers of massive ($> 10$\ M$_{\odot}$) progenitor stars.

It is important that the O-rich SNRs be studied as a class. 
An interesting and important as an object like Cas A is, it needs
to be viewed in the context of similar objects.
Each object contains distinctive features that allow us to investigate
different aspects of SNR dynamics and evolution to
attack fundamental questions concerning the origin of the elements: 
(1) Are these objects the result of Type II, Ib, or Ic supernova   
explosions? (2) Do the observational data from the            
various wavelengths lead to a consistent picture of SNR evolution?
(3) Do current emission and hydrodynamic models successfully            
account for the luminosities, morphologies, and kinematics of these 
objects? (4) Can theories of nucleosynthesis in massive stars     
and mixing in SN explosions explain the distribution of                
elemental abundances in the metal-rich ejecta?                    
(5) What are the probable progenitor stars of these SNRs, and are        
Wolf-Rayet stars viable candidates?
(6) Do the O-rich SNRs foreshadow the evolution of SN1987A?
(7) What will be the long-term evolution of these objects, and
how do they affect the chemical evolution of galaxies?

Important capabilities needed to make major progress in our
understanding of young SNRs in the Galaxy and Local Group
and the process of chemical enrichment of the ISM
are wide-field UV-optical imaging in diagnostic emission lines
and spatially resolved UV-optical spectroscopy.  
Such data can be combined with data obtained at other wavelengths,
such as X-ray and IR observations, to characterize the physical state
and abundances of the emitting gas.  Studies of SNRs and other nebular
objects require a selection of narrow-band filters that cover a
variety of ionization stages of several key elemental species. Ideally,
we would like to have filters in lines of H, He, C, N, O, Ne, Mg, Si,
S, Ar, Ca, Fe, and Ni. Such a large selection argues for the use of
tunable filters that would enable narrow-band imaging in any important
emission line over wide fields at arbitrary radial velocity/redshift
--- an important technology development issue. 

We must also study the dynamics and elemental abundance variations
within filaments, and from filament to filament, to test models of mixing.
Spatial resolutions $< 50$ mas are needed to isolate
specific filaments in Local Group SNRs and to resolve ionization
scale lengths in shocks. Fields of view $> 10'$ are desirable to map
the Galactic objects, including interesting jet-like protrusions that
suggest aymmetrical explosion geometries.
It is currently impossible with HST/STIS to obtain spatially resolved
UV spectra in the most important UV diagnostic lines (e.g., N~V, O~IV],
C~IV, Ne~IV], Mg~II) with spectral resolution high enough to compare
the UV kinematics directly to the motions deduced from optical 
emission-line profiles. Emission-line fluxes in the ejecta filaments 
are typically $F \approx 10^{-15} - 10^{-17}$ ergs cm$^{-2}$ s$^{-1}$
and require the next generation large-aperture UV-optical space 
telescope for further study.

\subsection{The Major Construction Phase of Galaxies and Quasars } 

\noindent
{\bf How and When were Galaxies Assembled?} \\ 

In the hierarchical clustering model of cosmological structure
formation, the rate of mass consumption from the IGM is tied directly
to the history of galaxy assembly and star formation.   
Both the merging of dark matter halos and the accretion of
small satellites determine the triggers to these phenomena.
Feedback from massive  star formation can also complicate the
astrophysical processes that govern radiative cooling and
compression of the baryon component within the dark-matter  
potentials.  These micro-physical processes depend sensitively
on the type of stars formed and on the initial mass function (IMF).

Over the history of the universe, massive stars dominated the
production of radiant energy and heavy elements, and the mechanical 
heating of the interstellar medium.  These O- and B-type stars
radiate the bulk of their luminosity in the rest-frame ultraviolet. 
The Galaxy Evolution Explorer (GALEX) will provide broad-band
spectral energy distributions of star-forming galaxies over the
range $0 < z < 2$, but will not obtain high-resolution spectra.  
The GALEX ultraviolet survey will provide the raw material for 
documenting the history of star formation and metal production 
during the crucial epoch from $z = 2 \rightarrow 0$, during 
which time the cosmic star-formation rate declined by more than an
order of magnitude. To understand the physical processes that 
drove the strong evolution during this era requires capabilities
for ultraviolet spectroscopy and high-resolution, wide-field 
optical imaging that greatly exceed those of HST.

First, UV spectroscopy with ST-2010 of galaxies in this redshift range
will measure both the IMF and the evolution in the chemical
composition of the massive stars. The same data will also
probe the gas-dynamical and chemical effects of the feedback provided
by the massive stars. These data will measure the rate at 
which star-forming galaxies injected metals and kinetic energy into 
the ISM and galactic halos in the form of galactic winds and fountains. 
An order-of-magnitude increase in UV throughput compared to HST/COS 
is required to reach the flux levels ($F_{\lambda} = 10^{-17}$
ergs cm$^{-2}$ s$^{-1}$ \AA$^{-1}$) and dispersions ($R = \lambda /
\Delta \lambda \approx 10^4$) needed to measure the stellar
composition.  In a 1 hr exposure of galaxies at $z = 1.7$, a 4m 
ST-2010 would be able to obtain a spectrum with S/N = 10 and 
resolution $R = 3000$ at the knee in the rest-frame, near-UV 
luminosity function (approximately 2.6 microjansky at 2900 \AA).  
ST-2010 could also perform spatially-resolved spectroscopy at
0.1 arcsec and beat out ground-based telescopes in the visible.
The typical half-light radius of the ``Lyman-dropout'' galaxies
is about 0.2 arcsec, so ST-2010 spectra of a few hours duration
could measure rotation curves, map regions of galactic outflows,
and measure spatial gradients of metallicity and age.   

Second, high-resolution, wide-field imaging in the 2000 -- 6000 \AA\ range
will document the cosmic evolution in the structure and
morphology of the young and intermediate stellar population in galaxies
during this epoch. With V-band imaging at high spatial resolution,
it will be possible to use population studies of nearby galaxies
to constrain the age of the populations, the timescales
for assembly, and the fraction of accreted material.
Among the specific topics to be addressed are: the separate
histories of bulge and disk formation; the roles of bars and galaxy
interactions/mergers in driving the star-formation rate and in
establishing the origins of the Hubble sequence and 
the density-morphology relationship for galaxies. Detailed 
observations of the stellar content and dynamics of local galaxies are 
just as crucial to understanding galaxy evolution. Constructing 
color-magnitude diagrams that reach the main-sequence turn-off in 
the nearest Local Group galaxies requires observations down to    
$m_B \approx 23-24$ (see Fig.~\ref{carina}).   Making such measurements 
of galaxies out to 3 Mpc is a major task, which is currently impossible 
for HST, even with ACS and WFC-3.  The ST-2010 imagers would provide the
best way of solving this problem. 

While the HST has made fundamental contributions in 
these areas, a complete understanding of how the modern universe of 
galaxies has emerged will require an increase of at least an order 
of magnitude in the field of view and a factor of three in imaging
sensitivity, compared to HST.  A large-aperture telescope with imaging 
and spectroscopy is required to perform the stellar population synthesis 
and spectral-line imaging to study the stars and gas at moderate redshifts.  
\begin{figure}[t]
\caption[]{The color-magnitude diagram for the Local Group dwarf
Carina at $D = 91$ kpc (Smecker-Hane et al.\ 1999). The main
sequence turnoff of one population occurs at $m_V \approx 23$, but
evidence for an older population occurs at $m_V \approx 24$.
Similar studies of more distant Local Group galaxies will require
imagers with wide field of view (FOV) and ten times more sensitivity 
than HST/ACS.  }
\label{carina}
\end{figure}

\vspace*{0.5cm}
\noindent
{\bf Wide-Field Imaging of Clusters and Superclusters} \\ 

The advantages of wide-field, optical imaging capability from a
space-based platform are substantial. While extensive studies
of large-scale structure at $z > 0.5$ will be common by 2010, the ability
to combine ground-based spectroscopic data with precision space-based
photometric and morphological information on galaxies within clusters,
filaments, and superclusters will still be limited by the relatively
small fields of view ($< 5'$) available in future HST and NGST
instruments.  The existence of superclusters at $z = 1$ is now confirmed. 
These systems span scales of $5-20h^{-1}$ Mpc. 
Here, $h$ is the Hubble constant in units of 100
km~s$^{-1}$ Mpc$^{-1}$ (current estimates suggest that $0.6 < h < 0.8$).
A $10-15'$ field of view (FOV) also provides a good 
match to the the correlation lengths of cosmological structures in galaxies 
and the IGM, to sizes of low-$z$ galaxy halos, and to quasar spacing 
on the sky at magnitudes $m_B = 19-20$.    

A thorough and efficient study of photometric redshifts in QSO
fields, as well of gradients in galaxy properties across clusters and 
superclusters, could be done with an imager that provides a 
$10' - 15'$ field of view. At $z=1$, a $10'$ FOV subtends $3h^{-1}$ Mpc, 
an order-of-magnitude gain over the discovery capability of ACS.   
Even clusters of galaxies, with sizes typically $1-3h^{-1}$ Mpc require
at least 4 ACS fields for proper study of the full dynamic range of the
intracluster environment. A $10'$ camera would allow such studies
to be done in a single pointing.  

Multi-channel, multi-object spectrographs (MC/MOS), soon to be
commissioned on 10m class ground-based telescopes, will have 
FOV that extend to $15'$ (e.g., the UCSC DEIMOS system for Keck).
As noted above, this is a good match to the QSO spacing at $m_V = 20$,
allowing a photometric survey of galaxies in nearly every QSO field.
Consequently, it is likely that many deep survey fields will be
studied spectroscopically by 2010. Ground-based photometry cannot 
provide accurate color measurements or precise morphological parameters
of galaxies lying much beyond $z = 0.2$, because of atmospheric seeing. 
The ability to
obtain efficient internal broad-band color and morphological data 
for an entire MC/MOS field in a single spacecraft pointing enables 
studies that would be extremely time-consuming with HST or NGST. 
One can use large-scale weak gravitational lensing (see Fig.~\ref{cluster}) 
to probe the underlying matter in galaxy clusters or galactic sheets.
The search for very rare objects, like
supernovae at $z > 1$, also requires wide-field, high resolution imaging.

By 2010, large, ground-based CCD mosaic imagers will be mapping
out several thousand square degrees of sky to faint limiting
magnitudes ($I > 23$).  Large-scale weak-lensing surveys based 
on such data will always be inherently limited by seeing effects;
one needs to measure ellipticities of faint ($I > 24$)
galaxies with high ($<10$\%) precision.  In general, the detection of 
very faint sources is much easier from space, where one avoids complexities 
of sky brightness and time-dependent point-spread
function gradients due to telescope flexure and atmospheric variations.
Given the present rate at which the area of CCD mosaic imagers have been
increasing in size (a factor of 25 over the past decade), it is reasonable to
expect that by 2010 a 16k $\times$ 16k system could be flown with few
technological challenges Even a 24k $\times$ 24k system might be possible.
Such CCD mosaics would provide angular resolutions of 25 to 40 mas, if a 
$10'$ field of view were achieved. This would enable diffraction-limited 
imaging at 4000 and 8000 \AA, respectively, for a space-based 4m and 8m
telescope. ST-2010 will probably out-perform NGST at blue wavelengths,
which provides the best match to the lensed galaxies. 

If a slitless spectroscopic mode is available with this imager 
(e.g., grism or prism), it might be possible to obtain redshifts for a large 
number of distant galaxies in a single pointing.  Photometric redshifts 
should be possible to $\Delta z \approx 0.02$ precision. 
The Lyman break, which lies in the UV for $z < 2.6$, provides a robust 
feature for estimating photometric redshifts, but it is currently limited
by the FOV and sensititivies on HST.  With medium-band filters
and a large-field near-UV imager, one could obtain photometric
redshifts for galaxies at $1 < z < 3$.  
A 4m class ST-2010 will enable more precise quantification 
of the intrinsic properties of $z < 1.5$ galaxies than HST and will
complement the work presumably to be done with NGST on $z > 3$ galaxies.
The advantages of an 8m class facility are that slitless 
spectroscopy at low-$z$ will rival and possibly exceed what can ever be 
achieved from ground-based high-redshift surveys, even from 10m class 
telescopes.  One would focus the slitless spectroscopy
on the range 2000 -- 4000 \AA, to study Ly$\alpha$ emitters
in the range $0.6 < z < 2.3$. At $z=1$, when the observed 
wavelength is 2432 \AA, the sky background is so low that detection
is extremely easy. With a $12'$ FOV, one can cover $3.6h^{-1}$ Mpc in
a single exposure, sufficient to map an entire cluster in a single
exposure or reach $20h^{-1}$ Mpc scales with a handful of adjacent exposures.
The flux of the Ly$\alpha$ line recently detected in the $z=5.6$
galaxy was $F \approx 10^{-17}$ ergs cm$^{-2}$ s$^{-1}$. 
Whereas it took 14,000 s on Keck/LRIS to obtain S/N = 10,
ST-2010 could obtain S/N = 10 (4m aperture) or S/N = 20 (8m aperture)
in only 2000 s. Surveys at $z < 2$ would nicely complement the work of 
NGST, which will focus on much higher redshift.   

\begin{figure}[t]
\caption[]{The distribution of dark matter in a galaxy cluster such as 
MS1137+6625 can be mapped using wide-field images of weak gravitational
lensing.  The dark matter distribution can be derived from
strong lensing in the cluster core evident in the HST/WFPC2 image.
But to map the dark matter over much larger scales via weak lensing
requires a larger field of view, like that obtained from ground-based
telescopes, and very high spatial resolution.
HST/WFPC2 image [left] courtesy of M. Donahue; R-band
ground-based image [right] from Luppino \& Gioia (1995).} 
\label{cluster}  
\end{figure}

\vspace*{0.5cm}
\noindent
{\bf Stellar Populations in Nearby Galaxies } \\
 
The stellar populations of nearby galaxies carry the fossil record of
galaxy formation.  The ages and abundances of the stars in these
galaxies constrain the timescales of their formation and assembly.
The rate of material accreted by these galaxies, as evidenced from the
distribution of stellar ages, also constrains cosmological models,
which predict definite and different accretion rates.  This fossil
record is difficult to decode from the integrated spectra and colors
of galaxies, particularly if NGST lacks the important V-band to
probe the flux peak of stars with the highest age sensitivity.
The decoding becomes progressively more accurate as individual
stars are resolved.  Historically, color-luminosity relations of
resolved stellar populations have been central to the development of
our astrophysical understanding of stellar and Galactic evolution. 
This can be expected to remain true even in the era of NGST and ST-2010.
 
Important features in the HR-diagram of old stellar populations include
the Main Sequence Turnoff (MSTO), the Horizontal Branch (HB), and the
Tip of the Red Giant Branch (TRGB).  The key to determining the age of
a stellar population is to measure the effective temperature of the
MSTO. That requires photometry accurate to $\pm 10$\% about 1 magnitude
below the turnoff.  Indirect constraints are possible via surface
brightness fluctuations in the near-UV (Worthey 1993).
At higher luminosities, the distribution of core-helium burning stars
on the HB and in the ``red clump," and the identification of RR-Lyrae
variables in particular, provides an indication of the existence of an
old population as well as an estimate of the chemical composition.
The ratio of the number of HB stars to the number of RGB stars
provides a measure of the helium abundance (Iben 1968).
At still higher luminosities, the colors and luminosities of
stars at the tip of the RGB constrain the possibilities for the age and
metallicity distribution of the underlying stellar population.

Information about the MSTO for Local Group galaxies is exceedingly
difficult to obtain (see Fig.~\ref{carina}). 
HST studies of the MSTO are now possible only for the Milky Way and
its immediate companions. After the installation of ACS it will be
possible with great effort to measure the MSTO in the halo of M31 and
M32.  The Horizontal Branch is accessible now with HST out to the
distance of M31.  The RGB tip is accessible with HST now out to the
distance of the Virgo Cluster.
A 4m class UV-optical telescope with better detectors would make
it possible to study the MSTO out to distances of 3 Mpc, bringing 
into view galaxies such as Centaurus~A, M81, M101, and their
complement of dwarf companions. The HB could be characterized out to the
distance of the Virgo Cluster, and the TRGB could be detected at the
distance of the Coma Cluster. With an 8-m class telescope the MSTO
of some of the nearest normal giant elliptical galaxies
(NGC~3379 for example) becomes accessible with great effort. SB
fluctuation studies in the NUV at the distance of Virgo would be
possible and would provide an estimate of the MSTO luminosity.
 
The temperature of the MSTO in even the oldest galaxies is $\sim5500$ K.  In
the absence of contamination from other types of stars, the V band is
the most efficient place to study the MSTO. In the inner regions of
galaxies, the light from the individual MSTO stars is overwhelmed by
that of neighboring RGB stars.  Observations shortward of 3000 \AA\ 
with strong red-leak suppression offer the only prospect for
isolating the MSTO near the centers of nearby galaxies such as M31.
A valuable output of the ST-2010 imagers would be a 
UV Spectral atlas of hot massive stars, a crucial ingredient in the 
templates used to interpret the UV spectra of distant star-forming galaxies.
 
\vspace{0.5cm}
\noindent
{\bf Cepheid Variables in the Coma Cluster} \\

Cepheid variable stars provide one of the primary distance indicators
used to establish the cosmic distance scale and Hubble expansion
rate ($H_0$) of the universe.  Although the MAP explorer hopes to 
determine $H_0$ to 10\% accuracy, this is comparable to the goals
of the HST Cepheid Key Project.  Even though some cosmologists believe
that $H_0$ and $\Omega_0$ will be measured conclusively during the next
decade, past history cautions us to be prudent and seek confirmation
of these critical cosmological parameters.  Both Cepheids and Type Ia
supernovae distance scales need to be reconciled with values
of $H_0$ determined from the microwave background fluctuations
and water-maser disk kinematics.  

Despite heroic efforts to reach the Centaurus Cluster, at 
distance modulus $\mu_0 = (m-M) = 33.0$, corresponding to 
distance $D = 40$ Mpc, by Zepf et al.\ (HST Program IDs \#6439 and \#7507), 
the practical limitation encountered by both the HST Key Project (Gibson et
al. 1999) and the Type Ia SNe Calibration Project (Saha et al.\ 1997) has
been $\mu_0 < 32.0$ (or $D < 25$ Mpc).  For the WFC, 
photon-starvation and crowding set a limit of $\sim 12$ pc~pixel$^{-1}$.   
With ST-2010, one could uncover Cepheid variables
directly at the distance of the Coma Cluster ($D = 85 \pm 10$ Mpc
or $\mu_0 = 34.65$).  Cepheids at Virgo have been discovered and monitored
by HST/WFPC at phase-weighted mean magnitudes $\langle m_V \rangle$ ranging
from 24.5 to 26.5.  At Coma, which is 5 times more distant
than Virgo, Cepheids would range from $\langle m_V \rangle = 28-30$.   
A high-resolution imager on a 4m ST-2010 would have three times the 
field of view of WFPC-2, and would draw upon the 3-fold increase in
collecting area over HST.  
It would also possess a resolution at Coma of 6 pc~pixel$^{-1}$.
 
The discovery of Cepheids in the Coma Cluster would allow a direct
determination of the Hubble Constant, via Cepheids, well out in the
Hubble Flow where peculiar velocities are $<$5\% of the recessional velocity. 
Other $H_0$ determinations, to date, generally require calibrating
other secondary indicators locally (out to Fornax) 
before pushing out into the Hubble Flow.  ST-2010 will
provide a unique opportunity to bypass this intermediate step,
yielding H$_0$ directly.  NGST will be non-optimal for
discovering Cepheids, because the light-curve amplitude is a factor of
two greater at V than it is at I.  Thus, Cepheid searches are almost 
exclusively attempted shortward of R.  Therefore, ST-2010 is the 
only facility on 
the horizon that will allow this direct probe of the cosmological
scale size, free from the effects of peculiar velocities that are
important for nearby galaxies.   
 
\vspace{0.5cm}
\noindent
{\bf Interactions of Galaxies with their Environment} \\

Perhaps the most important missing ingredient in our current understanding
of the formation and evolution of galaxies is the role played by
the back reaction of the energy produced by young stars on the
interstellar gas out of which the stars form and from which the galaxy 
is built.  This ``feedback'' takes the form of radiation and 
mechanical energy from massive stars and their evolutionary by-products. 
The tight correlations between such disparate galaxy properties as mass, 
size, surface brightness, velocity dispersion, and metallicity strongly 
suggest that galaxy-building is determined or regulated in some way by
this feedback. The simple cooling and dissipation of baryons in
dark matter potential wells does not by itself explain the properties of
present-day galaxies. In an even broader sense, the role played by the
radiation and kinetic/thermal energy supplied by massive stars may have
been central to determining the location, physical state, and chemical
composition of the baryons not presently incorporated into stars (which 
probably comprise the majority of baryons in the universe today).
 
Spectroscopy in the ultraviolet provides a unique suite of diagnostic
tools for investigating the physics of feedback and its astrophysical
consequences. The rich array of resonance absorption lines lying
in the rest-frame spectral region between the Lyman edge and roughly
2000 \AA\ makes it possible to study the dynamics, chemical 
abundances, and physical state of gas ranging from cold molecular
hydrogen to coronal-phase gas at $T > 10^5$ K.
By studying the gas seen in absorption against the UV continuum in
the star-forming regions themselves, we can directly study the process
of the feedback in action at its point of origin. These regions have
typical flux densities at 1400 \AA\ of $10^{-16}$ ergs cm$^{-2}$
s$^{-1}$ \AA$^{-1}$ and need to be studied at resolution $R \approx 
30,000$. Whereas HST/COS will only study typical starbursts at low 
resolution, $R \approx 3000$, a 4m ST-2010 mission could obtain
S/N = 10 spectra in several hours.  

Existing HST spectra
show that under the most extreme conditions (strong starburst nuclei),
cool and warm gas is being expelled at velocities of $10^2$ to $10^3$
km~s$^{-1}$, close to galaxy escape velocities, and at a rate
comparable to the star-formation rate. These data do not have high
enough spectral resolution or signal-to-noise to make more than
a qualitative characterization of the outflow, and then only for
a handful of the brightest starbursts.  An increase in spectral 
sensitivity by a factor of 10 would allow us to
undertake the necessarily detailed investigations of selected starburst
regions. Just as importantly, we could expand such investigations beyond
the extreme conditions in starbursts to more normal star-forming regions
in the disks of normal late-type galaxies. It would finally be possible
to investigate whether there is a threshold in
the rate of star-formation per unit area or volume above which
a galactic fountain or wind is established.
 
The cumulative effect of feedback and outflows can be traced by using
background quasars to probe the gas in galactic halos. Ironically,
this technique has taught us a lot about the properties of the gaseous
halos of galaxies at high redshift, while we are quite ignorant about
their counterparts in the local universe. This could change dramatically
in the next decade, if ST-2010 were available.

\vspace{0.5cm}
\noindent
{\bf Intervening Galaxy Halos} \\

As just noted, galaxy halos provide a transition or interface
between the high-density,
star-forming environment in galactic disks and the IGM. For galaxies 
at moderate redshift, their halos often extend beyond 100 kpc, and
can be studied in absorption against background QSOs.  Ultraviolet   
spectra are important in detecting and measuring physical parameters
of these halos, through resonance lines of ions such as C~IV, N~V,
O~VI, Si~III, Si~IV, and Mg~II.  As was done for the stellar winds of
hot, massive stars, UV spectra of galactic halos could detect galactic 
winds and outflows and quantify the chemical composition of the
flows injected into the IGM.  A spectral resolution $R \approx 10^4$
(30 km~s$^{-1}$) is needed to perform these abundance studies.   
 
Table 2.1 showed the anticipated gain in the number of quasars
and AGN if ST-2010 can reach $m_B = 19-20$.  These counts translate
into an effective average separation between objects, which will
allow ST-2010 to study sightlines toward various types of intervening 
galaxy halos from a ``structure of the universe" vantage point.  
At present, astronomers can only make spectroscopic observations of
a few halos.  At $m_V = 19$, ST-2010 will probably find one UV-bright
QSO every $20'-30'$, which can be used for absorption studies of many 
intervening galaxy disks and halos of different morphological types.
 
The Sloan Digital Sky Survey (SDSS) and the GALEX UV survey together
will measure redshifts, colors, and star-formation rates
for $\sim10^6$ galaxies in the ``local'' universe ($z < 0.2$) by 2005.
These surveys will also find several $\times 10^4$ UV-selected quasars 
brighter than $m_B = 18$. This implies that there will be roughly 500 
quasar-galaxy pairs with impact parameters (at the galaxy) less than 
about 100 kpc.  With a high-efficiency UV spectrograph, 
it will be possible to survey
this sample and determine the baryonic content, chemical abundances,
dynamics, and physical conditions in the halos of a sample of galaxies
spanning the complete manifold of galactic properties (mass, metallicity,
Hubble type, star-formation rate, starburst vs. normal galaxy, AGN host
galaxy vs. normal galaxy, cluster vs. group vs. field).
 
In addition, UV absorption lines seen in the lines of sight to luminous 
sources in nearby or distant galaxies permit the study of gas
in a number of important environments including the Milky Way halo and 
its system of high velocity clouds (HVCs), the halo gas  and HVCs 
of the target galaxy, and the environment associated with the luminous
source in the target galaxy.  The Milky Way is surrounded by a system of 
HVCs that likely have a number of origins.  Some of the possibilities
include gas pulled out of the Magellanic Clouds by tidal stripping, gas
ejected into the halo by vigorous galactic fountain activity in the
disk, remnant gas from the formation of the Milky Way, and Local Group
intergalactic gas.  Recent UV spectroscopic observations  with the HST
are now beginning to provide  important insights about the nature of the
HVCs.  Lu et al.\ (1998) have shown that the HVC in the direction of the AGN 
NGC~3783 has a metallicity that is most consistent with the HVC being
tidally stripped gas in the leading arm of the Magellanic stream.  Wakker 
et al.\ (1999) have studied UV and optical absorption toward Markarian~290,
which lies in the direction of Cloud Complex C, and find that this huge 
gas complex has a metallicity of approximately 0.09 times solar and is 
situated more than 3.5 kpc from the Galactic plane. They conclude that 
Complex C provides the first observational evidence for the accretion of 
low metallicity gas onto the Milky Way as required in current models of 
Galactic chemical evolution.  Sembach et al.\ (1999) have identified a new 
type of high velocity cloud (the highly ionized HVCs) which they believe 
may be associated with very low density and low pressure Local Group 
intergalactic gas.  

These interesting new results suggest that a 
full attack on the Milky Way HVC phenomena will provide a wealth of 
information about gas left over from the galaxy formation process,  
gas ejected into the halo by energetic phenomena occurring in the disk, 
and gas circulating around galaxies because of tidal interactions.  
To pursue a comprehensive study of the elemental abundances and physical 
conditions in the HVCs, it will be necessary to obtain high-resolution UV 
spectra of faint Galactic and extragalactic  sources situated beyond the  
HVCs.  While a few of the brighter suitably positioned extragalactic 
sources will be observed with STIS and COS on HST, a vigorous study of 
the HVC phenomena will require a spectroscopic facility more capable
by at least a factor of ten.  Spectral resolution of at least $R = 20,000$
is required to obtain accurate abundances. 

The observations required to study the HVC system of the Milky Way
could also be used to study the analogous phenomena associated with the
target galaxies.  The absorption at extragalactic velocities could be
used to answer such questions as: How common are systems of HVCs in other
galaxies?  Is there evidence for gas left over from the era of Galaxy
formation  in other systems of galaxies?  Do groups of galaxies have
associated clouds of intergalactic gas similar to the highly ionized
HVCs seen in the local group?  How do the abundances and physical
conditions in gas clouds observed at relatively low redshift, which 
are associated with current epoch galaxies, compare to those in clouds 
found in the high redshift universe observed through QSO absorption-line 
measurements?
 
The strongest UV emitting regions in external galaxies that could
be used as sources of continuua for the absorption line studies
discussed above are either associated with the luminous nuclear regions 
of the galaxies or with regions of enhanced star formation activity situated
elsewhere in the  disk of the external galaxy.  In such cases,
the UV spectra will  also yield direct information about the processes
whereby the starburst phenomena  releases vast amounts of mass, energy,
momentum, and chemically-enriched gas into the surrounding halos or 
intergalactic medium through supernova-driven galactic fountains or,
in the extreme cases, galactic winds.  Measures of the physical 
conditions and metal enrichment of this gas would provide information 
about the chemical enrichment and star-formation history of the galaxy 
and the extent to which starburst galaxies modify the chemical and 
physical environment of the surrounding halo gas or the surrounding IGM.

\vspace*{0.5cm}
\noindent
{\bf The Demographics of Massive Black Holes } \\
 
Massive black holes have long been suspected to be the central engines
of active galactic nuclei.  During the 1980s, this hypothesis moved closer
to proof with the finding that the central dynamics of a handful of
nearby galaxies could be explained by invoking a massive compact
object at their center.  
Given the exotic nature of black holes, however, most of the
initial work was heavily slanted towards finding more mundane
explanations for unusual central stellar dynamics.  The general
acceptance of massive black holes in galaxies was thus hard fought, and 
really only fully conceded for one or two systems.  Moving into the 1990s,
however, work with high-resolution spectrographs both on the ground and 
on HST, slowly fleshed out the picture that black holes indeed might be 
common to galaxy centers.
 
At the close of the century, the question has thus moved from the simple
question, ``Do black holes exist in galaxies?'' to richer issues of
``Are they a natural part of galaxies, and if so what role do they
play in their formation and evolution?''  Simple arguments imply
that black holes are ubiquitous. The integrated energy flux of QSOs over
the age of the universe, for example, suggests that fully 0.2\% of the
mass of galaxy spheroids may be in the form of black holes; today most
of this mass would be in the form of quiescent ``extinct,'' or more
accurately, ``dormant'' black holes at the centers of garden-variety
normal galaxies.  Further, because most galaxies observed
with sufficient central resolution do indeed appear to harbor central
black holes, the best hypothesis is that nearly all galaxies
have a central black hole.  In other words, the collection of fossil
black holes was not parceled out to just a few rare systems.
Indeed, the work to date suggests that central black mass correlates
with spheroid mass with a coefficient very close to that suggested
by QSO energetics.

Massive black holes thus may be the anchors around which galaxies form.
They may exert profound effects on the forms of central structure that
we see today, and their historical activity has perhaps moderated even
the global properties of forming galaxies.  Imaging work with HST, for
example, suggests that the central structure of the most massive galaxies
(characterized by low-density cores with shallow cusps)
can only be understood in terms of being built from the mergers
of less luminous galaxies (which have high-density steep central cusps)
with already extant black holes.  The latter systems themselves would
have been assembled around seed black holes formed in the early
universe.
 
The critical path to understanding the role of central black holes in
galaxies has thus evolved into an investigation of black hole
``demographics.''  At what stage in galaxy formation did a central
black hole form?  Do most galaxies indeed harbor central black holes?
Are they a critical part of all normal systems?  How are they parceled
out?  Is there a simple linear relationship between galaxy mass and 
central black holes mass?  Is the functional relationship more complex?
Are other parameters involved in the relationship such as galaxy type
or environment?  Given the answers to these questions, what has been
the back-effect of the central black hole on the structure and
dynamics of the rest of
the galaxy?  In the end, the goal is to unify the history of energetic
activity in the universe to formation of galaxies, with the link
between them being the massive black holes at the hearts of galaxies.
 
Work on black hole demographics has been started with HST, but progress
has been slow.  
By increasing the telescope aperture and detector sensitivity above 
that of HST, one expects to find many more black holess in more distant
galaxies.  Fixing the band luminosity, $\nu L_\nu$, for both 
quasar and host, the maximum redshift that can be studied increases 
as aperture to the 3/4 power.  Thus, doubling the mirror size allows 
one to move from the HST limit of $z \approx 0.3$ out to 
$z \approx 1.3$.  If the greater redshift increases the
luminosity in the observed band (likely to occur for both quasar and 
host), one does even better.  Therefore, a 4m mirror will take us from 
marginal studies of only the remnants of the quasar phenomenon, 
near the present epoch, to decent signal/noise at times close 
to the peak epoch of quasar activity.

The classic approach to studying central black holes
requires obtaining high spatial
resolution spectra at multiple points within a galaxy at sufficient
signal level to measure accurate line-of-sight velocity distributions. 
Unfortunately, slicing the central light into small pixels, 
all fed with a small mirror, biases the observations to galaxies
with the highest central surface brightnesses, which leads in turn
to strong biases in the global properties of the systems.
Imaging the centers of high-luminosity elliptical galaxies with low 
central surface brightness currently requires prohibitively long exposures. 
Observations of gas dynamics in galaxies helps reduce exposure time, 
when bright emission lines are found.  This introduces other 
biases in system selection. With the coming decade of HST observations, 
it may be possible to obtain a rough sketch of the relationship between 
galaxy luminosity and black hole mass over a significant luminosity range.  
However, it probably beyond the capability of HST to perform
an analysis of the width of the relationship and its general variability,
much less an investigation of additional parameters. 
With ST-2010, it is critical to invest sufficient time to probe systems 
in which the black hole may be non-existent or weak.  Basing a relationship 
on the galaxies that may be observed versus a systematic sample 
risks introducing profound biases in any picture of black hole demographics.

  \vspace*{0.5cm}
\noindent
{\bf Quasar Hosts } \\ 

The epoch $z \approx 2$ appears to corresponds both to the
greatest quasar activity and to the peak in cosmological star
formation.  Because the bulk of the starlight emerges in the
mid-ultraviolet (unless there is very strong dust extinction),
this light appears in our frame in the visible band.
The likely cutoff of NGST at 6000 \AA, or even 1 $\mu$m, 
places even more importance on ST-2010 for studies of the 
rest-frame UV of the galaxies.   

Perhaps the most fundamental unanswered question about active galactic
nuclei is why they exist at all.  Only the most speculative ideas
exist to explain why some galaxies create supermassive black holes in their
centers, and then feed them at a rate anywhere from 0.01 to 10 solar masses
per year.  Similarly, while it has been known for decades that the epoch
around redshift 2--3 was particularly conducive to the ignition of luminous
AGN, explanations of this fact are primitive at best. Perhaps it has
something to do with the much greater rate of galaxy encounters at that time;
or perhaps it was the result of the relatively large ratio of gas mass to 
stars that existed in galaxies then.  Detailed explanations are far beyond us.

ST-2010 could contribute to answering these questions in a number
of ways. First, there are indications that nearby AGN are associated 
with nuclear starbursts -- primarily an ultraviolet phenomenon.  Unfortunately,
HST is too small to permit spectroscopy of any but the brightest of these.
Installation of COS will not greatly change this situation, for it will
improve throughput by only a factor of a few in the mid-UV region that
is important for such studies, and it has no capacity for
spatially-resolved spectroscopy.  A larger throughput UV spectrograph with
either long slits or an integral-field system would permit detailed studies
of the stellar populations near the centers of local galaxies hosting AGN.
This might provide the conditions favorable to the creation of AGN, and 
possibly the time-relationships between AGN and starbursts.

Second, the studies that found starbursts in nearby AGN hosts used very 
optically thick dust obscuration located only $\sim 1$~pc to $\sim 100$~pc 
from the galactic nucleus as a ``coronagraph" (i.e., only type 2 Seyfert 
galaxies were observed).  Without that screening of 
the central nucleus, it would
have been much harder to observe the starlight in the inner several hundred
to 1 kpc of the host.  Unfortunately, very few obscured AGN are known
beyond redshifts $\sim 0.1$.  Consequently, the successful study of
quasar host galaxies depends critically on the ability to subtract 
(or artificially block) the light of the central nucleus.  This ability
is promoted by having a telescope with a larger aperture both because
of the gain in resolution (if it is diffraction-limited) and because of
the gain in signal/noise for observations having a fixed integration
time.  Rather than a conventional filled-aperture scheme, it might be 
better to use some sort of coronagraph in order to block out the quasar light.
Choosing the best size for the spot requires carefully thinking through
a trade-off:  Larger spots give better central source elimination, but also
cover more of the inner low surface brightness structure that is the object
of study.

Third, studies of quasar host galaxies during the epoch of greatest quasar 
activity ($z \approx 2$) should shed important light on triggers of 
AGN activity.  Through such studies, in concert with NGST data at longer
wavelengths, one can contrast galaxies {\it then} with galaxies {\it today}.
One can thus compare the mix of stellar ages and masses,
the chemical composition and physical conditions in the interstellar medium,
the presence of features like bars that might drive non-radial motions,
and the frequency of encounters.   
To be able to make these comparisons requires imaging with physical
resolution scale smaller than 1~kpc, i.e. $\sim 0.1^{\prime\prime}$ or less
for objects at cosmological distances.  Observations from space are clearly
required to meet this criterion alone.  The dark sky of space is also
important because of the $(1+z)^4$ dimming in bolometric surface brightness.

 Although HST has begun the study of quasar hosts, its results have been
more tantalizing than instructive (e.g., Bahcall et al.\ 1997; McLure et al.
1999).  For the reasons explained above, only a larger mirror will improve
this situation.  In fact, the effective signal/noise of these observations
is so sensitive to telescope aperture that doubling the size of the HST
mirror should move us from marginal results to data that can support
quantitative analysis.  ST2010 imaging will allow us to study the relationships
between the luminosity and other properties of the quasar and the luminosity and
morphology of its host.  ST2010 integral-field spectroscopy will extend
those investigations to quantitative determination of the host's stellar
population.  Astronomers have long speculated whether quasars are born
after their host galaxies are assembled, or ignition of a quasar
initiates formation of a galaxy around it.  With ST2010, we can hope
to answer questions such as this, and investigate more generally how
the two processes influence each other.
 
\vspace*{0.5cm}
\noindent
{\bf Gravitationally-Lensed Quasars} \\

     After twenty years of study, there are still only a few dozen confirmed
examples of gravitationally-lensed quasars.  Of these elements, 
a handful have been
particularly scientifically fruitful due to special circumstances.  For
example, the very first lensed quasar to be discovered, 0957+561, has
proven to be a good test-bed for detecting the inter-image time-lag;
the time-lag combined with measurement of the lensing galaxy's
potential has permitted an inference of the angular diameter distance
from us to the quasar.  Similarly, the Einstein Cross quasar (2237+0305)
is especially well-suited to observations of microlensing because the
lensing galaxy is so close to us.  As a result, the smooth lens inter-image
lag is very short, and the characteristic microlensing timescale is
compressed from decades to a few weeks.

     In the next few years, very large sky surveys like the Sloan Digital
Sky Survey should increase the number of known lensed quasars
tremendously.
Because the typical image separation is $\sim 1^{\prime\prime}$, imaging
from space will be essential for obtaining a clear picture of their
configurations, and instruments such as ACS should do a very good job
with their initial description.

     It is a reasonable expectation to suppose that some of the lensed
quasars to be discovered will have lensing properties even more favorable than
the examples already known.  From studies of these new quasars, we can
hope to obtain additional absolute distance measurements.  We can also
hope that there will be quasars in which monitoring of how the ultraviolet
flux varies during a microlensing event allows construction of an
``image" of the actual central engine.  However, we can also expect,
due to the sharp increase in the number of quasars with decreasing flux, that
many of these will be relatively faint.  Consequently, detailed study and
monitoring will require a telescope with better throughput, especially in the
ultraviolet, than HST.

\subsection{Other Scientific Programs}

There are numerous additional scientific programs
that would be impacted significantly by one or more future large-aperture
UV-optical space missions. Below we highlight the science goals
of several representative projects that would rely on the
high-throughput spectroscopic or wide-field imaging capabilities needed
to complete the core science mentioned above.
The investigations described do not comprise an exhaustive list of 
such projects but provide a flavor of the scientific diversity enabled
by a powerful facility for UV-optical astronomy from space.

\vspace{0.5cm}
\noindent
{\bf Origin of Stellar and Planetary Systems} \\
 
\noindent
{\it Protoplanetary disks.}
Stars form from dense molecular cloud cores which
are often so deeply embedded in their host clouds that they
cannot be directly investigated at visual wavelengths. However, in
many cases, such as when a nearby massive star ionizes the
surrounding medium, cloud cores and the young stars that form in them
can be exposed and rendered visible at UV and optical wavelengths.
Indeed, two of the most stunning results from HST are the discovery of
{\it proplyds} (proto-planetary disks)
in the Orion Nebula, which are young stars surrounded by
circumstellar disks embedded within the Orion Nebula that are ionized
from the outside by the hot Trapezium stars (e.g., Bally et al.\ 1998;
O'Dell \& Wong 1996), and the stunning
{\it ``Pillars of Creation''}, which consist of
isolated `elephant trunks' of dense molecular gas that have been
overrun by the expanding H~II region powered by a cluster of hot stars
in M16 (cf. Hester et al.\ 1996).  In both cases, the superior
resolving power of HST has provided new insights into star formation
and the structure of interstellar gas. However, it is only through the
laborious mosaicking of the Orion Nebula that a sufficient field
of view has been covered to study the entire star forming region.
The next frontier of star formation studies is to achieve wide
fields of view ($> 10'$) with high spatial resolution. The
HST/WFPC2 images show that it is necessary to obtain at least
50 mas resolution in order to trace the disk structure
and the disk-jet connection in proplyds in Orion, and a factor of
2 or better is desirable to extend these studies to the
many star-forming regions within 1--2.5 kpc of the Sun 
(Fig.~\ref{eagle}).
 
\begin{figure}
\caption[]{Evaporating gaseous globules (EGGs) that harbor nascent stars
are revealed in the HST images of the Eagle Nebula in M16 (top panels).
Studying the ``Big Picture'' of star formation and its feedback on
molecular clouds and the interstellar medium requires high-resolution
imaging over a much wider field, as shown in the ground-based image
(bottom panel).
HST/WFPC2 images from J. Hester, the WFPC2 IDT, and NASA; ground-based
image obtained at KPNO courtesy of J. Bally.}
\label{eagle}
\end{figure}

The vast majority ($> 90$\%) of stars currently forming within $\sim 2.5$ kpc
of the Sun form in Orion-type environments. If planetary systems are common in
the Galaxy, they must be able to form and survive within these surroundings.
Studies must therefore show that (a) proto-planetary disks are common
in OB associations, and (b) planet formation can occur in the face of
energetic processes (e.g., photoevaporation, stellar winds, supernovae)
that destroy disks. Questions we might address include, How does planet
formation depend on the IMF and massive stellar content? How does disk
survival depend on distance from the massive stars? What are the disk
properties as a function of position within the star-forming regions?
Clearly, IR and sub-mm observations will contribute significantly to
answering such questions; however, considerable important and unique
information about proto-planetary disks can be gathered from optical
observations as well.

Observational programs to be carried out with the next generation UVO
space telescope may include: (1) A census of circumstellar disks
in major star-forming regions within 2.5 kpc of the Sun. Optical imaging
from space of dark disks surrounding nascent stars against the bright 
background emission of ionized gas is very efficient for finding and studying
proto-planetary disk systems at sub-Solar System scales 
(e.g., Bally et al.\ 1998; Fig.~\ref{hst10}). Extending these studies
to distant H~II regions requires high spatial resolution and wide-field
imaging. Such a census could be combined with IR studies of embedded
sources in nearby clouds that have not yet been exposed by the presence
of massive stars in the vicinity. (2) Study of the physics of disk/planet
formation and survival in star-forming environments. The high angular
resolution of space imaging allows us to probe important physical scales
in diagnostic UVO tracers of gaseous flows and mass loss. In particular,
is the timescale to form $\sim 1$ cm size bodies in circumstellar disks
that are resistant to photoevaporation shorter than the disk destruction
timescale? (3) Resolve the relationship between massive stellar content 
and the IMF/star formation efficiency. For example, do massive stars 
trigger or hinder solar-type star formation in surrounding clouds?

\begin{figure}[h]
\caption{Protoplanetary disks in Orion are seen against the background
emission in this optical HST-WFPC2 image provided by J. Bally.
The object near the center is clearly being disrupted by the energetic winds
and photon field of nearby massive stars. Such processes are predicted to
destroy the disks on relatively short timescales. The object in the
upper-left, on the other hand, appears to be a foreground star/disk system
which is evidently outside the influence of the destructive forces
of the massive stars. High-resolution imaging of such disks
against the bright background emission could reveal the presence of gaps
in the disks that may indicate planet formation has occurred.}
\label{hst10}
\end{figure}
 
\noindent
{\it Protostellar outflows.}
Our current picture of star formation is that during gravitational
collapse, angular momentum inherited from the parent
cloud core results in the formation of a disk through
which much of the mass is accreted onto the young stellar object (YSO).
It is believed that magnetic fields regulate the accretion rate,
the dissipation of angular momentum, and the ejection of powerful
outflows along the rotation axis of the system.
The earliest phases ($\leq$ 2 Myr) in the evolution of
proto-planetary/accretion
disks are best probed by indirect observations of the jets and
outflows that YSOs produce. The spatial distribution,
morphology, and velocity field of shock excited gas in such
outflows preserve a fossil record of the sequence of mass ejection
events that are driven by stellar accretion, over a time-scale
comparable to the formation time of a young star.
The outflows from young stars manifest themselves as
(1) loosely collimated bipolar molecular
flows which carry large amounts of energy
and mass, or (2) highly collimated stellar jets that move at speeds
of several hundred km~s$^{-1}$ away from the star and become visible
as material cools behind shock waves in the flow.
The optically emitting radiative shock waves
are called Herbig-Haro (HH) objects, and evidence is mounting
that the high-velocity outflows drive the molecular outflows
through `prompt entrainment' processes at the jet/ISM interface
(e.g., Heathcote et al.~1996; Reipurth et al.~1997).
 
The past several years have seen significant advances
in our understanding of protostellar outflows:
(1) Ground-based wide-field, narrow-band CCD imaging of several well-known
star forming regions in Taurus, Perseus, and Orion has revealed hundreds
of new Herbig-Haro outflows, supporting the notion that all stars
undergo a phase of energetic mass loss during formation.
These (optical) flows typically span parsec scales and may range up
to 10 pc in total length.  It is now clear that protostellar
outflows impact their local molecular clouds over much larger
spatial scales and longer time scales than previously believed.
(2) Stellar jets are highly variable, both in mass flux, which generates
a series of bow shocks, and in ejection direction, so that even very
highly collimated outflows are able to excavate large cavities in the
surrounding
cloud.  (3) The high spatial resolution of HST has
finally resolved individual shocks in protostellar jets.
For the first time, we can see how shocks propagate from
the jet to the surrounding medium,
a process that transfers energy and momentum to molecular outflows.
The structure and kinematics of protostellar jets
provide clues to the conditions close to the YSO which
are inaccessible by direct observation.  The challenge is to decipher
the information encoded in the jet structure and draw
conclusions about the accretion process.
 
Future research will address the physics of the shock waves in
protostellar flows, the extent (and nature) of the entrainment of 
surrounding material,
and the cumulative impact the outflows have on
star forming regions and the interstellar medium (ISM).
In particular, we need to resolve the shock structures in jets and
monitor their propagation, cooling timescales, and directional
variability. These goals require narrow-band imaging of
Galactic star-forming regions in diagnostic emission lines over
$> 10'$ fields with better than 50 mas (few AU) spatial resolution
at multiple epochs. We also need to extend our studies to major
star-forming regions in the Local group, such as the dynamic 30 Dor
region in the LMC (cf. Scowen et al.\ 1998). Studying such local
`mini-starbursts' bears directly on our understanding of massive star
formation in young galaxies at high redshift.

\vspace{0.5cm}
\noindent
{\bf UV Absorption Lines in our Galaxy: Stars with Heavy Extinction} \\
 
At ultraviolet wavelengths, stars dim very rapidly when extra
interstellar material is added in front.  For the usual gas-to-dust
ratio and extinction law for dense clouds in our Galaxy, the logarithm
of the flux at 1150 \AA\ decreases at a rate $-6.4\times 10^{-22}{\rm
cm}^2N({\rm H_{total}})$, relative to its value without the obscuration.
This rapid attenuation for stars inside or behind dense clouds makes
them very hard to observe in the UV, but at the same time it makes them
more interesting to study.  As a result of the shielding of dissociating
UV radiation by dust grains and molecular hydrogen absorption features,
molecules can survive for long times in the cloud interiors and, as a
consequence, lead to profound changes in the chemical makeup.  This
simple conclusion is confirmed by detailed theoretical models and
observations at infrared and sub-mm wavelengths that show that molecules
are indeed plentiful at the higher extinctions.  For instance,
observations of the $3.05~\mu$m ice-band absorption feature indicate
that ice-coated grains have appreciable concentrations only within the
densest portions of compact clouds.  For such clouds, the approximate
linear trends in $\tau(3.05~\mu{\rm m})$ vs. $A_{\rm V}$ extrapolate to
zero ice absorption for $A_{\rm V}$ that ranges from 2.6 to 5.0 in
different surveys.
 
While radio and IR techniques can probe many kinds of molecules in the
gas phase, their sensitivities are significantly worse than those
achievable from UV absorption studies.  It follows that research on most
molecular constituents can, as a rule, only be conducted for very dense
regions.  In instances where the optical depths become large, studies of
emission features in the radio region are confounded by radiation
transfer effects.  Completely out of reach of the IR and sub-mm
observations are the single atoms and ions -- these constituents can
only be studied at UV wavelengths (and to a very limited extent, in the
visible).  Unfortunately, limitations in present-day telescope and
spectrograph sensitivities have prevented us from doing research on
stars with visual extinctions in excess of $A_{\rm V}\approx 1$.  In
short, the important regime where the interstellar medium undergoes a
transition from pure atomic to pure molecular states has been an elusive one.

It should be possible to observe clouds with $A_{\rm V}$ up to 10 or so
with a 4--5m class UV telescope.  This limit may even be exceeded, but
this depends critically on how the UV extinction varies with wavelength.
If dust grains continue a trend seen in some clouds and stick together
to make a significant number of larger grain clusters, the extinction
law could become favorable enough to permit us to probe the centers of
clouds that are about to form new stars.  To study such clouds in the UV
would give important insights on the depletions of atomic constituents
and the concentrations of various molecules in the gas phase at
different depths within the clouds.  Comparisons with observations of IR
solid-state features along the same sight lines should yield important
clues on the affinity for certain molecules to the surfaces of dust
grains.


\newpage  
\noindent
{\bf UV Absorption Lines in our Galaxy: Halo Stars of Moderate Luminosity} \\
 
With HST, we can probe the gaseous matter in the halo of the Galaxy by
observing extragalactic objects that are bright in the UV (quasars and
Seyfert nuclei) or rare, main-sequence, early-type stars that are at
large distances from the Galactic plane.  The problem with both kinds of
objects is that there are not very many of them that are bright enough
to study with much precision in the UV.  The brightest stars in the
Magellanic Clouds are of some use in principle, but the picture they
present may be misleading due to interference from tidally stripped
gas along the line of sight.  
Extragalactic sources have the shortcoming that they are of no
use in determining distances of the intervening material, as one might
hope to learn from stars within the Galactic halo.  Thus, the best that
we can do at present is to conduct a very sparse study of the halo, and
we could reach misleading conclusions because we are studying material
that is very patchy.
 
With a more powerful telescope, we can overcome the problem of sparse
sampling. Stars on the blue end of the horizontal branch that have
effective temperatures in excess of 20,000~K are plentiful, and apart
from their faintness they are excellent light sources in the UV.  Such
stars at distances ranging from 5 to 40 kpc from the plane have visual
magnitudes between 13.5 and 19.5, and they can offer sweeping insights
on the nature of the High Velocity Clouds, the distribution of
coronal-type gas ($10^5~K < T < 10^6$~K), the driving forces of a
``Galactic Fountain'' (if one exists), how well the gas follows 
galactic rotation as a function of distance from the plane.  This 
information, in turn, will tell us how to interpret the absorption 
lines in quasar spectra that are produced by gases in the halos of 
other galaxies.

\vspace{0.5cm}
\noindent  
{\bf Planetary Science: Origin and Nature of Solar Systems} \\ 
 
The discovery of extra-solar planets is historic, opening new
areas of research that overlap with classical astronomy, offering
new targets for study, and changing the way planetary scientists
think. At the same time, we need to first characterize the planets
in our solar system to be able to understand extra-solar planets.
At the August 1998 workshop, two main questions emerged from the
solar system panel discussion, which are entirely consistent with
the scientific goals of the NASA Strategic Plan:
 
\begin{itemize}
 
\item How and why do planetary systems form, and what do they look like
when they do?
 
\item What makes planets evolve into habitable worlds?
 
\end{itemize}
 
Today an increasing fraction of planetary science is being done by
remote observations, and given the high cost and difficulty in
sending missions to the more distant planets, we expect that
the planets and satellites far from the Sun will continue to
be studied mainly by remote observations.  By virtue of their
distance from the Sun, these objects are also the ones most likely
to be detected around other nearby stars.  The advent of a
new mission, ST-2010, with a discovery efficiency approxmately 50 times
that expected for HST/COS, will make available important new
measurements of distant planets and satellites in our solar
system.  
 
\vspace*{0.5cm}
\noindent
{\bf The Nature of Distant Planetary Surfaces and Atmospheres} \\
 
  Solar reflectivity spectra of planets and satellites give
critical compositional information about their atmospheres and
surfaces, and UV spectra at wavelengths below 2000 {\AA} sample
the most important and sensitive transitions of simple atoms
and molecules.  The present HST sensitivity limits spectra of
distant and/or small objects (e.g., Galilean and more distant
satellites, Neptune, Pluto) to near-UV wavelengths above
roughly 2000 {\AA}.  The ST-2010 increase in effective area will
extend UV spectra to more distant and fainter objects, including
asteroids and possibly Kuiper belt objects, and extend
wavelengths down to the critical UV range where simple atoms
and molecules have the strongest absorptions.  Another promising
technique for studying planetary and satellite atmospheres is
to make long-aperture spatially resolved emission scale height
measurements, from a variety of atmospheres and for different
emission lines.
 
\vspace*{0.5cm}
\noindent
{\bf High-Resolution Studies of Planetary 
Atmospheres by Stellar Occultations} \\
 
  Visible and UV stellar occultations provide altitude profiles
of the atmospheric composition of planetary and satellite
atmospheres, with altitude resolution proportional to the time
resolution and thereby signal to noise.  UV occultations
furthermore provide the highest sensitivity to small columns of
planetary upper atmospheres and satellite atmospheres (e.g., Io,
Ganymede, Triton, Pluto etc.).  However, the present rate of
suitable candidate events is 1 per several years with HST. 
ST-2010 will increase the effective area for spectroscopy by
a factor of 10 over HST/COS.  This will greatly
increase the number of occultations available and the signal to
noise of each event.  UV observations are critical, especially when
taken in conjunction with groundbased visible or IR occultation
observations.  Some expected benefits of UV occultations
include the detection of haze or clouds, estimates of optical depth
due to haze or clouds, and the determination of atmospheric
constituents (e.g., CH$_4$, from its ionization cutoff).
One example of a situation where UV observations would have been
useful is the 1988 stellar occultation by Pluto. A kink in the
occultation lightcurve at a Pluto radius of 1215 km may be due to
(a) a haze layer, (b) a thermal inversion, or (c) some combination
thereof.  UV observations of this occultation by a spectrograph
with the capability of ST-2010 would have resolved this basic question.  

\vspace*{0.5cm}
\noindent
{\bf Direct Detection and Characterization of Extra-Solar Planets} \\
 
The present detections of extra-solar planets proceed from stellar
radial velocity measurements.  Direct imaging in the thermal
IR is planned for NGST, where the contrast of planetary to stellar
emission is expected to be highest.  At the same time, there are
several candidate observations in the UV which may prove to be
sensitive indicators of the composition of both mature and forming
planetary systems.   One is the direct detection and
characterization of the known extra-solar planets, many of
which are Jupiter-sized and located near the star.   
During an occultation of a star by the planet, the 
planet's extended atmosphere can be studied by observing the
time-variable absorption with wavelength within the broad stellar
emission line profiles.  High sensitivity UV spectra of known disks
and planets would be obtained spatially resolved from
the parent star to characterize the composition and/or disk temporal
absorptions from infalling material (as in $\beta$ Pic).  Direct
imaging would reveal far-UV emissions from planet-forming material
near young, nearby stars.  Finally, high-sensitivity UV spectra of
known brown dwarfs, either isolated or spatially resolved from
companion star, would characterize their composition and
excitation properties.  It may be some years before the true
potential of these observations is known, but these examples
illustrate some of the most exciting science that
might be accomplished by a future NASA UV mission.

\vspace*{0.5cm}
\noindent
{\bf Stellar Outflows across the H-R Diagram } \\

For stellar ejecta, the primary science goal is to probe the structure of  
outflows down to a few thousand stellar radii or smaller in order to 
understand how the winds and outflows are formed, accelerated, and,    
apparently, collimated into highly structured patterns.
Even though many types of stellar outflows are nonisotropic, they do have
a high degree of point or reflection symmetry.  This says that the process
that governs the outflow is anchored to the star one way or another.  The
most common classes of physical models with this attribute are magnetic
fields, rotationally-driven processes of one sort or another, or complex
combinations of the two as in T-Tauris, for example.

Models that might explain the nonisotropic outflows are at a very early
stage of their evolution.  They desperately need more geometric and
kinematic data for constraints.  The classes of models divide into two main
groups: processes that work within a few stellar radii (e.g.,
wind-compressed flows and disks), and those that rely on the ambient
molecular medium (such as a torus) for a structured pressure environment.
The size scales are thus about $10^{12}$ cm ($10^{-4}$ arcsec at 1 kpc) 
or about $10^{16}$ cm (1 arcsec at 1 kpc).
Observations seem to take little heed of these models, however.  WFPC2
observations of several types of collimated outflows (e.g., bipolar PNe and
H-H objects) show structure on every visible scale ($10^{15}$ cm or larger).
It seems plausible that the models have adopted the wrong paradigms.

Progress will require both the best possible spatial resolution and
dynamic range so that the near-stellar flow region can be probed.  This
translates into the need for filled, large, obstruction-free apertures,
perhaps with an off-axis secondary.  Of high (but secondary) importance is
the need for excellent spectral coverage (from Paschen $\alpha$ at 1.87
microns) to N V] in the mid-UV.  A tunable filter is ideal; however it must
be able to reject nearby bright lines separated by 15~\AA\ 
(e.g., [N~II]~6548 and H$\alpha$; H$\gamma$ and [O~III] 4363; 
[S~II] 6717/6731) with a relative
transmission of $10^{-4}$.  It might be useful to aim for resolving the
[O~II] 3736/3729 and C~III] 1907/1909 lines for diagnostic purposes.


\section{UV-OPTICAL MISSION CONCEPTS FOR THE POST-HST ERA}

There is consensus on the UVOWG that any substantial hiatus
between the de-orbit of HST and launch of the next generation UV-optical
space astronomy mission(s) would severely compromise crucial aspects of our
ability to study astrophysical phenomena. The exquisite imaging and
versatile spectroscopic capabilities of HST have provided a myriad
of important discoveries during the 1990s that have sparked a
fervor among researchers as well as the general public. With the addition 
of new high-throughput imaging and spectroscopic instruments (ACS, WFC3,
COS), HST will continue to produce forefront scientific results that arise
from its uniqueness, even with the advent of numerous 8-meter class
ground-based telescopes during the next decade. In its recommendation of
the next generation space telescope concepts for the post-HST era, the UVOWG
refers to ``ST-2010'' missions to emphasize the need to sustain forefront
UV-optical capabilities from space, along with the need to press forward
on technology development programs that would enable
significant new scientific studies not possible with current missions.

We identify mission options to undertake the science described in the
previous sections. Depending on the pace of technology development and
its impact on mission costs, one may regard these options as a roadmap.
We present performance requirements for mission concepts with 4-meter (Class I)
and 8-meter (Class II) apertures. The Class I missions are designed to
accomplish directed science goals, to obtain ``discovery factors'' roughly an
order of magnitude larger than current capabilities, and to be cost-constrained
to SIRTF-class experiments. The Class II mission will achieve tremendous
gains in sensitivity, though may depend on technological developments that
are not feasible in the next decade. The Class II mission will draw on
heritage from other space missions by leveraging technology investments,
such as in the area of large deployable mirrors, but also relies on the
development of new detector technologies that will enhance mission performance.
Ultimately, each mission concept should be capable of pursuing a broad range
of science objectives by achieving UV-optical sensitivity limits comparable
to or exceeding the largest ground-based telescopes and providing key
complementary capabilities to NGST.

When contemplating the viability and impact of various mission concepts,
we must consider the following areas for trades and technology
assessment/development.
\begin{enumerate}
\item {\bf System throughput:} Geometric aperture size, number of
optics, spectral efficiency/coatings, grating efficiency, detector
quantum efficiency.
\item {\bf Wavelength coverage:} Coatings, detectors, number of
observing modes.
\item {\bf Telescope configuration:} Field of view, resolution, wavefront
error, tolerancing, facility size.
\item {\bf Large primary mirror configuration:}  Substrate material;
Monolith vs.\ segmented; Launched in-place vs.\ on-orbit deployment;
Segment shape, size, figure, polish, coating; Testing.
\item {\bf Telescope alignment and testing:} Wavefront
sensing/correction, active correction on primary mirror vs.\ on
secondary or internal to instruments.
\item {\bf Spectrometer configurations:} Number of optics/modes,
resolution, wavelength coverage, detector format and type.
\item {\bf Imaging camera configurations:} Field of view, resolution,
wavelength coverage/filter selection, detector format and type.
\item {\bf Orbit:} LEO, HEO (elliptical orbit), geosynchronous, or L2.
\end{enumerate}

Figures of merit can be derived that relate the performance of a mission
to current capabilities, especially those aboard HST. In order to
accomplish the science goals described in Sec.\ 2, we must achieve substantial
improvements in OTA+instrument performance, which we term the
``discovery factor'' (similar to its use to compare the performance 
of ACS to WFPC2).  For spectroscopy, the discovery factor ${\rm F_S}$
is a combination of throughput (or effective area, ${\rm A_{eff}}$) and simultaneous
wavelength coverage (${\rm F_S = A_{eff} \times \lambda_{range}}$) for a given
spectral resolution needed to complete the science goals. 
In the case of a multi-object or integral field spectrograph, we also
account for the spatial multiplexing capability. For imaging, the
discovery factor ${\rm F_I}$ combines throughput and field of view
(${\rm F_I = A_{eff} \times FOV}$) for a given spatial resolution needed to
complete the imaging science goals, assumed to be superior to that afforded
over wide fields by ground-based telescopes or Explorer-class
missions at the wavelengths observed.
We also can increase the observing efficiency, ${\rm O_{eff}}$, compared to
HST by choosing an orbit that allows long continuous target visibility periods
and that lowers or eliminates bright-Earth and geocoronal emissions that
contribute to background count rates. Combining the discovery factors with
observing efficiency, we can create metrics called discovery efficiencies:
${\rm D_S = F_S \times O_{eff}}$ for spectroscopy, and
${\rm D_I = F_I \times O_{eff}}$ for imaging.

Table 3.1 summarizes some of the performance characteristics of the
Class I and II missions that are described in more detail below.
\begin{table}[ht]
\begin{center}
Table 3.1: Class I and II Mission Performance Compared to HST \\
\  \\
\scriptsize
\begin{tabular}{|c|c|c|c|c|c|l|l|}
\tableline
Mission & Aperture & Mode & $\lambda$ & Resolution & Field &
Discovery & Discovery \\
        &          &      & Coverage  &            & of View &
Factor  & Efficiency \\
\tableline
\tableline
           &       &              &                 &                 &
           &                      &                        \\
ST-2010    & 4.2-m & Point-source & 1150 -- 3200 \AA & R $\geq$ 30,000 &
$\sim 2''$ & ${\rm F_S \approx 50}$ & ${\rm D_S \approx 100}$ \\
Class Ia   &       & Spectroscopy &                 &                 &
           &                      &                        \\
           &       &              &                 &                 &
           &                      &                        \\
\tableline
           &       &              &                 &                 &
           &                      &                        \\
ST-2010    & 4.2-m & High-Res & 0.2 -- 1 $\mu$m & 30 mas @5000 \AA &
$4\farcm1 \times 4\farcm1$ & ${\rm F_I \approx 5}$ & ${\rm D_I \approx 10}$ \\
Class Ib   &   & 16k$\times$16k Imager &             & (15 mas pixel$^{-1}$) &
           &                      &                        \\
           &       &              &                 &                 &
           &                      &                        \\
           &       & Wide-Field & 0.2 -- 1 $\mu$m & 30 mas @5000 \AA &
$13\farcm6\times13\farcm6$ & ${\rm F_I \approx 50}$ & ${\rm D_I \approx 100}$ \\
           &   & 16k$\times$16k Imager &             & (50 mas pixel$^{-1}$) &
           &                      &                        \\
           &       &              &                 &                 &
           &                      &                        \\
           &       & Integral   & 0.35 -- 1 $\mu$m & 30 mas $\mu$-lens$^{-1}$&
$8'' \times 8''$ & ${\rm F_S \approx 240}$ & ${\rm D_S \approx 480}$ \\
           &       & Field Spect. &             & R $\approx$ 5000 -- 10,000 &
           &                      &                        \\
           &       &              &                 &                 &
           &                      &                        \\
\tableline
           &       &              &                 &                 &
           &                      &                        \\
Class II   & 8-m   & Point-source & 1150 -- 3200 \AA & R $\geq$ 30,000 &
$\sim 2''$ & ${\rm F_S \approx 250}$ & ${\rm D_S \approx 500}$ \\
           &       & Spectroscopy &                 & (1-D STJ array) &
           &                      &                        \\
           &       &              &                 &                 &
           &                      &                        \\
           &       & High-Res & 0.2 -- 1 $\mu$m & 15 mas @5000 \AA &
$3\farcm3 \times 3\farcm3$ & ${\rm F_I \approx 10}$ & ${\rm D_I \approx 20}$ \\
           &    & 24k$\times$24k Imager &             & (8 mas pixel$^{-1}$) &
           &                      &                        \\
           &       &              &                 &                 &
           &                      &                        \\
           &       & Wide-Field & 0.2 -- 1 $\mu$m & 15 mas @5000 \AA &
$12\farcm3\times12\farcm3$&${\rm F_I \approx 150}$ & ${\rm D_I \approx 300}$ \\
           &    & 24k$\times$24k Imager &             & (30 mas pixel$^{-1}$) &
           &                      &                        \\
           &       &              &                 &                 &
           &                      &                        \\
           &       & Integral   & 0.35 -- 1 $\mu$m & 15 mas $\mu$-lens$^{-1}$&
$8'' \times 8''$ & ${\rm F_S \approx 880}$ & ${\rm D_S \approx 1760}$ \\
           &       & Field Spect. &             & R $\approx$ 5000 -- 10,000 &
           &                      &                        \\
           &       &              &                 &                 &
           &                      &                        \\
\tableline
\end{tabular}
\end{center}
\end{table}

\subsection{Class I Mission Concepts}

The Class I mission concepts are of $\sim$ 4-meter aperture.
There are three principal drivers behind pursuing this size aperture:
\begin{enumerate}
\item {\it Scientific return} --- Factors of 10 to several hundred gain
in discovery efficiencies are possible over current capabilities,
enabling frontier scientific studies not currently possible.
\item {\it Technological feasibility} --- There is a clear technological path
to implementing missions of this class, both in terms of mirror development
and advancements in instrumentation.  A 4.2-meter monolithic mirror can just
fit inside a Delta-class launch vehicle (i.e., segmented mirror technology is
not needed). Assuming NGST weight constraints on the
total mass, we can scale the 12 kg/m$^2$ restriction for an 8-meter
NGST primary mirror to $\sim$44 kg/m$^2$ for the ST-2010 primary mirror.
This latter performance has already been met by the secondary
mirrors manufactured for the VLT telescopes.
\item {\it Cost constraints} --- Extrapolation of several proposed Discovery
missions of 2.4-m aperture up to a 4-m class aperture suggests that a 4-m
mission is feasible within a SIRTF-class cost envelope ($\sim$ \$220M
for the telescope, $\sim$ \$50M for an efficient spectrometer,
$\sim$ \$100M for a very wide-field imager, and \$60-80M for launch).
\end{enumerate}

\subsubsection{ST-2010 Class Ia Mission (4-m)}

    The Ia mission is a 4-m class observatory optimized for UV
point-source spectroscopy. This concept may be viewed as the most
affordable option, as the main science goals do not require
diffraction-limited imaging nor wide field of view.
However, as a UV optimized mission, the requirements on mirror surface
roughness and scattered light are still likely to be stringent.
The UV point-source spectroscopy science goals drive the following
mission characteristics.

\begin{enumerate}
\item{Orbit:} geosynchronous or L2 orbit
\begin{itemize}
\item Achieve high observing efficiency                                
\item Operate simply and efficiently
\item Achieve low sky backgrounds for long-duration exposures
\begin{itemize}                                                        
\item Minimize or eliminate geocoronal emission                     
\item Shield or discriminate against cosmic rays
\end{itemize}                                                   
\end{itemize}
\item{Spectral Resolution:}
\begin{itemize}
\item Optimized for (slitless) point-source spectroscopy
\item Spectral resolution R = 30,000 -- 50,000 for primary science
\item Faint-object ``survey mode'' (R = 1000)
\end{itemize}
\item{Effective Area:} At least $10\times$ HST/COS
\begin{itemize}
\item Achieve ${\rm A_{eff}} > 2 \times 10^4$ cm$^2$ 
(target AB = 17-20 mag QSOs)
\item 4.2-meter primary aperture
\item Next generation (low-background, high-QE) detectors
\item High-efficiency UV coatings; holographic gratings; minimal reflections
\end{itemize}
\item{Spectral Multiplexing:}
\begin{itemize}
\item No more than 2 integrations to cover 1150-3200 \AA\ at R = 30,000
\end{itemize}
\item{Spatial Resolution:}
\begin{itemize}                                                    
\item 0.3 arcsec or better (Rayleigh criterion)
\end{itemize}
\item{Wavelength Coverage:}
\begin{itemize}                                                    
\item Current HST UV band: 1150-3200 \AA
\end{itemize}
\item{Minimal Imaging:}
\begin{itemize}                                                    
\item Off-the-shelf CCD imager for tracking/target acquisition
\end{itemize}
\item{Launch Date:} 2010
\begin{itemize}                                                    
\item Ready when HST is de-orbited
\end{itemize}
\item{Mission Lifetime:}
\begin{itemize}                                                    
\item Design lifetime 5 yrs; mission goal 10 yrs
\end{itemize}
\end{enumerate}

A large increase in the discovery efficiency of the Ia mission compared
to HST will depend on several technological developments (see Sec.\ 4
for more details):
\begin{itemize}
\item Flight qualification of lightweight 4.2-m monolith mirror
\item 2- or 3-bounce optical design of slitless UV spectrograph
that compensates for figure control and aberrations
\item Next generation of large-format MCP-based detectors with low dark count 
($< 0.1$ cts s$^{-1}$ cm$^{-2}$) and high-QE performance ($> 60$\%) over a
broad wavelength range
\item High-reflectivity, broad-band UV coatings
\item High-efficiency ($> 60$\%) holographic gratings
\end{itemize}

The effective area of the HST/COS R=20,000 spectral modes is predicted to be 
${\rm A_{eff} \approx 1500-2000}$ cm$^2$ in the FUV and half that in the NUV.
A factor of 10 gain in effective area over HST/COS must arise from a
combination of larger aperture, higher OTA UV reflectivities, more efficient
gratings, and higher QE detectors. We achieve a
factor of $\sim$3 gain in collecting area with a 4.2-meter telescope
compared to HST. It is estimated that the HST OTA delivers roughly 50\%
of incident Ly$\alpha$ photons to the focal plane, implying reflectivities
of the primary and secondary mirrors of $\sim$70\%. OTA reflectivities of
$\sim$85\% have been achieved with the COSTAR and STIS optics, and so could
supply a gain of a factor of nearly 1.5 over HST. The COS (1st-order)
gratings have groove efficiencies of about 60\%, and the next generation
holographic gratings may deliver as high as 75\% (see Sec.\ 4), providing 
a factor of 1.25 gain. These three improvements combine to give a factor
of $\sim$5.6 gain in effective area over HST/COS. Hence, to achieve a factor
of 10 gain requires that the detector QE improve by almost a factor of two,
in the FUV from $\sim$30\% to $\sim$60\%. Clearly, the largest burden is on
achieving higher QE detectors.

The next component of the discovery factor is wavelength coverage.
The COS R = 20,000 FUV modes deliver $\sim$300 \AA\ coverage and the NUV
modes $\sim$150 \AA\ using 1st-order gratings.
Covering the entire 1150 -- 3200 \AA\ range
in two integrations requires either much larger detector formats and/or much
better detector resolution. Alternatively, we could use echelle gratings,
as is done on STIS.  This approach may compromise our groove
efficiencies and add an additional optic, hence not aid our net gain in
discovery factor. Assuming, however, that we can solve the problem of
providing broad wavelength coverage with efficient gratings and large
detectors gains us a factor of $\sim$5 advantage over HST/COS, so that the
total discovery factor could be as high as ${\rm F_S \approx 50}$.

It will take a concerted technology development effort to achieve
a ten-fold increase in effective area over HST/COS, although the
issues appear tractable.  When we multiply by the factor of $\sim 2$
greater observing efficiency by operating in a geosynchronous or L2 orbit,
we attain a gain in discovery efficiency of ${\rm D_S \approx 100}$ over
HST/COS.  Such gains provide the capability to attack the point-source
spectroscopy science goals detailed in Sec.\ 2.

\subsubsection{ST-2010 Class Ib Mission (4-m)}

Every image of the sky that HST takes tells us something new about the
Universe in which we live. This continues to be true even for bright objects
such as the Orion nebula that have been studied for literally hundreds
of years. The new discoveries generally derive from the improved spatial
resolution and sensitivity that HST delivers versus ground-based imaging,
which enables us to investigate important scale lengths in faint
structures from the planetary arena to cosmological scales. HST's
advantage in spatial resolution has declined to only about a factor of
2-3 in recent years with the advent of adaptive optics (AO) systems on
ground-based telescopes (cf. the AO bonnette system at CFHT). With
progress on ground-based and space-based optical interferometers over
the next decade, we can expect many experiments to supersede HST's
spatial resolution, in some cases by orders of magnitude.  However, neither
AO-assisted imaging nor optical interferometry will achieve these high
resolutions over wide fields of view with high dynamic range,
and, of course, imaging in the UV is
completely unique to space-based astronomy.

When considering the next frontier of UV-optical imaging from space,
there are several avenues to explore, such as spatial resolution,
field of view, limiting sensitivity, and dynamic range.  The science goals
of, e.g., mapping the distribution of dark matter in superclusters,
conducting UV imaging surveys of nearby galaxies, and studying the
origin of stellar and planetary systems in Galactic and Local Group
star-forming regions all require the ability to detect faint targets
distributed essentially randomly over large areas on the sky with spatial
resolution high enough to discern important physical scales.

The ST-2010 Mission Ib 4-m class concept is designed to accomplish these
performance goals, occupying an extremely important and unique region of
parameter space that is rich in discovery potential.
The essential characteristics of the Ib mission follow.
These are in addition to many of the mission properties identified in
the Ia mission, such as orbit, launch date, mission lifetime,
lightweight mirrors, and high-reflectivity coatings, that are shared by
the Ib mission.

\noindent{Wide-field NUV-Optical Imaging:}
\begin{itemize}
\item 0.2-1 $\mu$m wavelength coverage (overlapping with NGST in at least one
band in the red)
\item Broad-band and narrow-band imaging
\item Sky coverage greater than 10$\times$ HST/ACS Wide Field Channel
\item Full point spread function (PSF) correction for diffraction-limited
performance at 5000 \AA\ (= 30 mas), matching the resolution of an 8-m NGST
diffraction-limited at 1 micron
\end{itemize}

Critically sampling with 15 mas pixels a field of view (FOV) that covers
10 times more sky than the Wide Field Channel of ACS would require a CCD
detector array roughly 44k $\times$ 44k in size, perhaps unrealistically large.
However, it may not be unrealistic to pursue arrays 16k $\times$ 16k, or
perhaps even 24k $\times$ 24k in size, as large-format mosaics of this
scope are currently being constructed for ground-based cameras.
If we baseline a 16k $\times$ 16k CCD, critically sampling the FOV
with 15 mas pixels yields a $\sim 4' \times 4'$ field, congruent with
the baseline FOV performance of NGST. Using a pixel size of 50 mas yields a
field over $13' \times 13'$, 16 times larger than the FOV of the ACS WFC
and commensurate with the imaging science requirements detailed in Sec.\ 2.
This pixel scale would undersample the PSF, however, the Hubble Deep
Field images made with 100 mas pixels clearly show that dithering
techniques recover information and yield sufficient resolution to
produce frontier science. Implementing both pixel scales to provide 
a high-resolution, critically sampled mode as well as a wide-field survey
mode could be accomplished either by sharing the focal plane of the
telescope and feeding two separate CCD mosaics, or by means of a focal
reducer that alters the plate scale delivered to a single detector mosaic.

The ST-2010 survey imaging mode with 50 mas pixels has the same
pixel scale as the ACS WFC. With a 4.2-m aperture, the ST-2010 collecting area
provides a factor of $\sim$3 greater sensitivity over HST/ACS. Then even
with very modest improvements in detector QE and surface reflectivities,
the discovery factor is 50 times higher than HST/ACS. Operating in
high-Earth orbit --- provided a proper solution to cosmic ray shielding
can be implemented --- may increase the observing efficiency by as much
as two times, leading to a discovery efficiency enhancement for the
ST-2010 wide field imaging mission Ib of ${\rm D_I \sim 100}$.
Larger enhancements may be obtained in the NUV (0.2 -- 0.3 microns) because
current CCD QE performance at these wavelengths is substantially lower
than in the visible bandpass.

Science goals such as exploring the kinematics and physical
conditions in galaxy cores and supermassive black holes require 
spatially resolved spectroscopy, and we can take advantage of the
excellent image quality of the Ib mission to provide this capability.
The simplest application may be a long-slit mode, though achieving
complete spatial sampling with an integral field spectrograph would be
more desirable for such investigations. Spectral resolutions of 
R $\sim$ 5000 -- 10,000 are adequate for most kinematic studies and
resolve important diagnostic absorption and emission lines.
It would be desirable to use the existing large-format detector to capture
the spectra from a micro-lens array at least a couple hundred elements
on a side, in order to provide fine spatial sampling (30 mas) over a
field $\sim 8'' \times 8''$ in size.
The relevant discovery efficiency comparison is with HST/STIS, which
would require stepping a long-slit across the FOV in order to obtain
complete spatial sampling. Comparing to the available 0\farcs1-wide
STIS slit, a huge gain in discovery factor (${\rm F_S \approx 240}$) 
results from the complete spatial sampling of the integral field unit and
the increased collecting area of the telescope. Including the gain in
observing efficiency of operating in high-Earth orbit yields a discovery
efficiency of ${\rm D_S \approx 480}$.

Achieving the performance of the Class Ib mission requires the
development of several technological capabilities.
\begin{enumerate}
\item Develop lightweight 4-m class monolithic primary mirror
with excellent figure and/or (active) figure control.
\item Optical design that achieves wide FOV and high spatial resolution.
\item Requires excellent tracking and pointing stability.
\item Develop large CCD mosaics for space.
\item Improve detector QE performance, especially in the near-UV.
\item Lower read noise and dark currents.
\item Improve charge transfer efficiency performance.
\item Implement shielding against harmful cosmic rays.
\item Develop UV-optical tunable filters for broad-band and narrow-band imaging,
that provide access to numerous diagnostic lines at arbitrary redshift.
\end{enumerate}

\subsubsection{Stretch Goals}
The following stretch goals would enhance the performance of the Class I
missions and increase the scientific return if such capabilities can be
made both feasible and affordable.
\begin{enumerate}
\item Achieve spectral resolution of R = 50,000 -- 200,000.
\item Extend UV imaging down to 1150 \AA\ to include access to rest
Ly$\alpha$, C IV, and other diagnostic lines.
\item Extend spectroscopy to the FUSE band (912 -- 1180 \AA).
\item Achieve diffraction-limited imaging in the UV ($< 10$ mas resolution).
\item Include UV-optical coronagraphic mode for high contrast imaging studies.
\end{enumerate}

\subsubsection{Combining the Class Ia and Ib Missions}

Integrating the Ia and Ib missions together into a single mission could be
accomplished by sharing the focal plane between the imaging and point-source
spectroscopy channels (similar to the way HST shares its focal plane
among multiple instruments). The point-source spectroscopy channel does not
require a field of view more than a few arcseconds, which could be accommodated
at the periphery of the imaging field.  Technological advances that constrain
mission costs are very important in this concept in order to fit this more
capable, higher science return concept within the SIRTF-class cost envelope.
Figure~\ref{concepts_fig}
shows an optical design concept, based on a scalable NGST design.
In this example, the imaging and spatially resolved spectroscopy channel
with full PSF correction is fed by an off-axis tertiary, while the on-axis
beam feeds directly into a point-source UV spectrograph.
Many other design concepts are possible.

\begin{figure}[ht]
\caption{Optical design concept for a large-aperture UV-optical
space telescope that shares the focal plane between wide-field
imaging and spectroscopy instruments. This optical layout, with the
fast steering and PSF correction mechanisms housed in the instrument
module, borrows heavily from the yardstick NGST design concept.
Many other design concepts are possible.
(Contributed by Dr.\ Charles Lillie, TRW.)}
\label{concepts_fig}
\end{figure}

\subsection{Class II Mission Concept (8-m)}

The Class II 8-m mission offers tremendous gains in discovery efficiency
over current capabilities.  This option would provide enormous scientific
return, and would maximize the return on NGST technology investments.
While the scope of this mission may delay its readiness beyond the
HST de-orbit, there is a logical path towards its implementation.
Significant funds are being invested in developing technologies to
build a passively cooled 8-meter telescope for NGST. For those requirements in
common, it may be less expensive to utilize NGST developments and designs
directly for a UV-optical mission, rather than develop independent technology.
Depending on the technology selections made for NGST, common requirements
could include optical surface supports, structures, deployment mechanisms,
spacecraft elements, actuators, control systems, and power systems.  Better
surface accuracy is required for the UV-optical than for the IR, but the
optical surfaces would not be cooled, so the temperature gradients and
accompanying structural distortions would be lower. Adaptive optics controls
would require less range, and so could be more accurate.

If an 8-m UV-optical mission were envisioned by NASA, we must
take full advantage of the large aperture by employing efficient
instrumentation with the very latest detector technologies.  For example,
an efficient spectrograph would follow the same strategies as for the 
Class Ia telescope option, a minimum of surfaces and high-efficiency,
low-scatter gratings.  A resolving power of 20,000 -- 30,000 can be
obtained with a single curved echelle grating, with one exposure over the
912 -- 3000 \AA\ range. An energy resolving detector, such as a
superconducting tunnel junction (STJ) device, with 50 -- 100 resolving
power could sort the orders and separate them by more than 3 FWHM over that
range. Alternatively, a prism could order-sort for a non-energy resolving
two-dimensional detector.

Some of the essential characteristics of the Class II mission follow.
\begin{itemize}
\item 8-m class telescope in L2 orbit that leverages NGST segmented mirror
technology investment and on-orbit deployment and operations experience
\item Utilize NGST-type packaging into launch vehicle
\item Some performance goals are more difficult than
for NGST: e.g, diffraction-limited
performance in optical or UV; co-phasing of mirror segments
\item Some performance goals are easier: e.g., cold telescope not necessary;
relaxed thermal requirements and fewer pointing restrictions
\item Employ energy-resolving, high-QE (STJ-type) detectors to maximize
performance and efficiency
\end{itemize}

The large 8-m aperture represents a significant increase in collecting
area over HST. When combined with an efficient spectrograph that employs
high-QE ($\sim 80$\%) detectors, it is possible to obtain an effective
area of ${\rm A_{eff} \sim 1 \times 10^5}$ cm$^2$. Obtaining the
complete UV spectrum in one exposure using a curved echelle grating and
operating in high-Earth orbit may yield a gain in discovery efficiency
of ${\rm D_I \approx 500}$ compared to HST/COS!

Ostensibly, a large-format optical camera could be included for very
deep, wide-field imaging. However, the prospects for achieving
diffraction-limited imaging ($\sim$15 mas at 5000 \AA\ for an 8-m telescope)
in the optical with the segmented mirror technology are more dubious at this
time compared to the Class I monolithic missions. A 24k $\times$ 24k
CCD detector array with 30 mas pixels could reach faint limiting magnitudes
eleven times faster than HST/ACS and with 13 times greater sky coverage.

Some basic development issues for the Class II mission (in addition to
those mentioned for the Class I missions) are:
\begin{enumerate}
\item Achieve diffraction-limited imaging at optical wavelengths with segmented
8-m mirror design, leveraging NGST mirror technology development and
on-orbit experience.
\item Develop innovative optical design to accommodate both high-resolution
spectroscopy and wide-field imaging (probably not an NGST ``clone'').
\item Development of concave echelle gratings that can be employed in a
Rowland circle-type, high-throughput spectrograph.
\item Develop large (1-d or 2-d array) energy-resolving detectors.
Only small arrays ($\sim 6 \times 6$) currently exist; arrays of at least
$1 \times 2048$ (1-d, spectroscopy only) or 2k $\times$ 2k (imaging and
spectroscopy) are needed.
Point-source echelle spectroscopy: 1-d detector array with sufficient
energy resolution (R = 50 -- 100) to sort orders.
Imaging and long-slit spectroscopy: 2-d detector array large enough for
significant FOV and with sufficient energy resolution (R $\approx$ 200 -- 500)
to obtain crude redshifts and for simultaneous narrow-band imaging in numerous
UV-optical diagnostic lines.
\item STJs operate at milli-K (!) temperatures, requiring space qualification
of next generation cryogenic technology.
\item Pointing stability may be difficult, and a fast tip-tilt seconday may
be necessary. But for a point-source spectroscopy mission, good stability is
only necessary in the dispersion direction.
\end{enumerate}

Finally, we briefly mention an intriguing possibility of flying a
descoped implementation of the ``chord-fold'' NGST design without the
chords --- i.e., a 4m $\times$ 8m elliptical telescope that would fit into
a Delta-class launch vehicle. The PSF may not be suitable for detailed
imaging studies, but would serve as a ``light bucket'' for point-source
spectroscopy.

\subsection{Pathfinder Mission}

We briefly mention the possibility of devising a pathfinder mission
for high-throughput UV spectroscopy. The roadmap to very large
($>$ 20m) aperture space telescopes of the future will require the development
of ultra-light optics, such as thin film deployable mirrors. 
A 10-m class pathfinder mission using the new thin film technology 
would yield ground-breaking sensitivity limits for UV spectroscopy, while
providing a platform for testing the deployment, image quality control,
and operations of a large aperture telescope.

\subsection{Additional Missions}

The UVOWG was charged with considering the next frontier of UV-optical
space astronomy in the post-HST era. We did not spend a large amount of
time considering small, Explorer-class mission concepts, for which NASA
already has an implementation process. However, there were new Explorer
and Discovery class UV-optical mission concepts presented at the Boulder 
conference with science goals that the UVOWG endorse. An underlying
theme of several of these mission concepts is to provide dedicated
facilities to study time-variable phenomena, following the important 
science executed by IUE during the latter years of its mission life.
We also provide a brief description of a UV interferometer concept 
that would provide ultra-high spatial resolution.

\begin{enumerate}
\item Explorer class UV spectroscopic mission dedicated to long-term  
monitoring of time variable sources, such as cataclysmic variables,
young stellar objects, and active galactic nuclei. Time-resolved spectroscopic
data will be used to diagnose the physical conditions in accreting
systems and to study the disk/jet connections so ubiquitous in astrophysics.

\item Explorer/Discovery class optical photometry mission dedicated to
detecting extra-solar planets via occultations of the central stars
by planetary bodies (especially those in 51 Peg-type orbits), or via
microlensing events as distant stars are lensed by planetary systems
orbiting foreground stars.

\item Discovery class UV-optical telescope dedicated to remote
sensing (imaging and spectroscopy) of time-variable phenomena
occuring in planets in the Solar System, as well as characterization of
extra-solar planets (e.g., 51 Peg-type systems) via changes
in the central star spectrum due to occultations by the extended
atmospheres of giant planets as a function of orbital phase.

\item Explorer class mission dedicated to imaging and spectropolarimetry.
Polarimetric observations provide a unique and powerful technique for
determining the three-dimensional structure and geometry of many
astrophysical objects.  Polarization is also useful for mapping magnetic
fields at scales ranging from the stellar to planetary and as a means of
determining the characteristics or thermodynamic properties of interstellar,
proto-planetary, or cometary gas and dust.  Observations of polarization are
particularly useful in the UV, with access to strong resonance lines
for determining scattering geometries, physical conditions, and
composition in numerous astrophysical objects.  The primary
caveat is that polarimetric measurements require high S/N to achieve meaningful
results, and the instrumentation requires elaborate and precise calibration.
The current suite of sub-orbital experiments (with 5-10 minute
missions) have been limited to a small list of bright targets.  Building
from the existing instrumentation, which have honed the techniques, the
next logical step is the development of an Explorer-class polarimetric
observatory.  Even a modest ($\sim$0.3m) aperture instrument would dramatically
increase the number of accessible objects and provide an unequivocal
demonstration of the power of the technique.

\item Ultra-high resolution imaging with a space UV interferometer.
Science goals include (1) resolving surface features on nearby stars;
(2) obtaining $\geq 10$ resolution
elements across the inner 0.1 AU region of the
nearest protostars to image hot gas in accretion flux tubes to address
how stars are made;
(3) resolving 1 AU (or better) scales in the Magellanic Clouds
for studying stellar systems and interstellar processes
along a sight-line with low reddening; (4) obtaining $\geq 10$
resolution elements over 0.1 pc scales in the nearest AGN to study
galaxy cores and black holes;
(5) resolving 1-10 pc scales at 1 Gpc distance to study individual
star-forming groups in high-redshift galaxies.
\end{enumerate}

\newpage

\section{TECHNOLOGY ROADMAP}

\subsection{Overview}

Throughput is the single most important technology driver for 
the future of UV-optical space astronomy, especially for spectroscopy.
Significant astrophysical problems as discussed above --- the 
formation and early evolution of galaxies, the nature of dark matter,
the formation and (re)cycling of elements, the nature of the dynamic 
interstellar/intergalactic medium, the formation and early evolution 
of galaxies --- cannot be properly addressed now because of the lack of 
sensitivity to low-surface-brightness or intrinsically faint objects.   
Throughput can be improved significantly by (1) using more sensitive 
detectors, (2) using significantly larger aperture telescopes, and 
(3) by improving the efficiency of the instruments, especially through 
clever designs and improved optical surfaces.  Throughput can also be 
improved greatly by achieving high levels of multiplexing, 
particularly in spectroscopic applications.

UV-optical studies of faint or low-surface-brightness objects also 
require low backgrounds (natural or instrumental), good signal-to-noise 
over a wide range of signal strengths, and linear responses.
These goals can be addressed by using (1) low-noise 
detectors with (2) high dynamic range.   

Technology needed to advance UV-Optical astronomy can and will 
benefit greatly from advances being made in other wavelength regimes.
In particular, detector developments in the X-ray, ground-based visible,
and infrared wavelength bands are invaluable and can often serve 
as good starting points for UV detector development. Large, lightweight 
optics are of interest to the infrared/visible community (e.g., NGST), 
where investments in optical materials and deployment mechanisms 
are already showing positive results.  

However, the UV/space visible regime is unique in several respects.
In the UV (10 -- 300 nm), the potential for contamination and subsequent
drastic loss of throughput must be considered in every part of an 
instrument and spacecraft.   The short wavelengths of UV light make 
it intrinsically more difficult (than at visible or near-IR wavelengths)
to produce precision optics, to align those optics, and to maintain 
the optical alignment and wavefront in large optics.  In some cases, 
the diagnostically crucial UV emissions are weak relative to many other 
regions of the electromagnetic spectrum --- thus detectors that 
are sensitive to UV light may be much {\it more} sensitive to (and perhaps 
overwhelmed by) the much brighter red/infrared emission from 
astrophysical objects.

Because of these unique constraints, there are several areas where 
UV technology development cannot rely on advancements made
in other wavelength regimes, but will progress only through dedicated specific
efforts.  These areas include:
\begin{itemize}
\item Detectors
\item Large lightweight precision mirrors
\item Optical materials and coatings
\item Precision optical elements --- gratings, micro-mirrors
\end{itemize}

A chart illustrating the flow down of the science requirements to the
ST-2010 telescope/instrument performance and technology development is
shown in Fig.~\ref{flow_chart}.

\begin{figure}
\caption[]{Flow down of ST-2010 science requirements to
telescope/instrument performance and technology development.
(Prepared by Randy Kimble.)}
\label{flow_chart}
\end{figure}

We wish to emphasize that major breakthroughs in technology development
for UV-optical space astronomy, especially in the area of detectors, 
almost certainly require a more robust technology development initiative
from NASA than is currently in place. For example, the recently closed
NRA 98-OSS-10 on ``Technology Development For NASA Explorer Missions
and SOFIA'' was heavily over-subscribed, extremely broad-based in its
technology response, and funded at a level that is generally 
insufficient to provide sustained development of major
new hardware over years of effort. Many programs find it necessary 
to bootstrap multiple sources of funding in order to sustain 
instrument development.  We encourage NASA to supplement current
technology and instrumentation programs to take full advantage
of community efforts to develop new concepts or improve existing
technologies that, in the long run, result in large mission cost savings.

\subsection{Detectors}
 
\subsubsection{Overarching Requirements}

The primary science drivers for the future of UV-optical astrophysics
require medium- to high-resolution UV spectroscopy and wide-field
UV-optical imaging of faint extragalactic targets.
These science goals place demanding requirements on the 
sensitivity of the detector systems. High quantum efficiency (QE)
and low background levels are the critical parameters for the detector(s),
both for spectroscopic and imaging applications, and they warrant the greatest 
investments in detector technology.   Highly efficient multiplexing 
systems should be developed for use at UV-optical wavelengths, with 
particular attention paid to 3-D (energy-resolving) detectors.
Work should also be done to obtain larger formats, to improve dynamic 
range (critical for precise calibration), and to increase stability 
(for high S/N operation). 

{\it High quantum efficiency} --- All promising lines of significant QE 
improvement should be explored.  Without major gains in instrument 
sensitivity, many of the observations required to advance the field of 
UV-optical astronomy cannot be done. UV astronomy has the dubious distinction 
that there is a great deal of room for improvement.  The most commonly used 
UV detectors, MCP-based photon counters, provide QEs 
that are far from unity (typically 10-40\%, depending on wavelength).
Detector QE offers a great deal of potential leverage on instrumental
performance. {\it A given factor of improvement in QE provides as much return
in sensitivity as a corresponding increase in telescope aperture, 
at potentially much lower cost.}

{\it Low backgrounds} ---  It is essential to reduce substantially the 
background rates of all types of detectors.  UV-optical studies of faint, 
diffuse objects require low backgrounds to detect the signal from 
astrophysical objects.  In long observations of the faintest targets, 
detector background noise becomes the limiting factor and determines the 
sensitivity of the measurement.  Accumulated background counts from the 
detector (``dark count'') may overwhelm the signal from the target object,
and the purely statistical fluctuations in the background counts produce a 
fundamental noise floor on top of which target counts must be measured.
For CCD-like detectors, electrical fluctuations in the readout of each 
exposure (``read noise'') add an even larger component of detector 
background noise.

{\it Dynamic range and linearity} --- Technology investments should be made
to ensure that potential flight detectors have large dynamic ranges, and are 
stable and linear.   It is essential to be able to observe fields containing 
bright sources without saturating or damaging the detector.   It is also 
important when exploring scientific problems that involve small 
perturbations in the signals from very bright sources, that detectors provide
a linear response to the signals.  Finally, it is crucial, in making such 
measurements, that detector signal-to-noise (S/N) ratios are not compromised 
by unpredictable position-dependent (and thus uncalibratable) irregularities 
in detector response. 

{\it Multiplexing} --- Accomplishing the science goals outlined in the
previous sections depends on achieving large, simultaneous spectral
wavelength coverage. Large (2-D) detector formats are needed to meet
this requirement. We may also look to developing ``3-D'' detectors
to improve efficiency and reduce the number of optics. We suggest
that a major program of development for energy-resolving
UV-optical detectors should be undertaken.  Sensitive, multidimensional
detectors with both good spatial and moderate-to-good spectral resolution
are particularly important.  This technology requires an extensive and 
sustained development program. Overall efficiency of an instrument can also 
be improved by increasing the number of objects observed simultaneously.
This can be accomplished by using larger fields (in imaging applications),
or by obtaining multiple spectra simultaneously, or ideally by doing both.
Promising avenues for improving UV-optical multiplexing beyond the 3-D 
detectors include development of micromirror arrays or of UV-transmitting
optical fibers. 

In the following sections, we comment on the state of the art and 
developments required in several specific detector types.

\subsubsection{Multidimensional Detectors}

The so-called 3-D detectors, such as superconducting tunnel junction 
(STJ) devices and transition edge sensors (TES), have the potential to 
revolutionize UV-optical astrophysics.   With significant and sustained 
development work, these detectors could be ready to fly in the timeframe
of the missions described above.
Substantial, long-term, stable funding should be invested to bring
these or related 3-D detectors to the astronomy community.

3-D detectors provide photon-counting with intrinsic energy resolution of each 
detected event.  While the energy resolutions currently envisioned are coarse,
compared with the spectral resolution requirements presented above, it should
be possible to use a linear array of such energy-resolving pixels to provide
order-sorting in a high-resolution echelle spectrograph, thereby eliminating
the need for a cross-disperser grating and providing a high QE over a
wide simultaneous wavelength coverage. A large 2-dimensional array of these
devices would provide extraordinarily efficient multi-object spectroscopy.  

Serious engineering issues remain to be addressed before these detectors become 
practical. STJ/TES detectors require cryogenic operations, with UV-optical
applications expected to require sub-Kelvin temperatures and the requisite 
coolers, dewars, and windows. It is likely that the requisite coolers will be 
developed to support other wavebands (e.g., the sub-mm).  However, many of the 
other cryogenic requirements must be explored and integrated into standard 
thinking about UV-optical instrumentation.  Current prototype arrays are only 
$\sim 6 \times 6$ pixels (e.g., see Jakobsen 1999; Perryman et al.\ 1999).
Techniques must be developed to create and read out large arrays; then 
the fabrication of such arrays must be developed.  Current versions of 
these detectors are also quite sensitive to red and near-IR light; 
future development should stress utilizing this broad-band
capability or implementing blocking-filter concepts to reduce red sensitivity 
while maintaining the excellent UV response, depending on specific application.

\subsubsection{Semiconductive Arrays}

CCDs have brought about revolutionary increases in capability to record 
information at visible wavelengths, because of their high QE and their linear 
responses over a large dynamic range.  In the ultraviolet, CCDs are less 
attractive, because of their much lower QEs as well as their 
excellent response to unwanted visible/red light.  New developments are making
these devices more attractive for UV work, particularly for near-UV applications
where the ``red leak'' is not as important (e.g., spectroscopy of intrinsically
blue QSOs. Future work should concentrate on reducing read noise, 
utilizing alternate materials, developing photon-counting versions of these
detectors, and implementing very large format detector arrays for space-based
applications.

{\it Silicon-based CCDs} --- CCD detectors have recently made substantial
gains in UV response, currently offering QEs significantly higher than those 
of photo-emissive detectors at NUV wavelengths above 200 nm
(e.g., see Clampin 1999).  However, for long, background-limited observations
of faint sources, even a substantial QE advantage for a CCD detector will be 
overwhelmed by the read-noise penalty of repeated reads (required for cosmic 
ray rejection).  Until CCD read-noise can be brought well under 
1 e$^-$ pixel$^{-1}$ rms, CCDs will not compete successfully with MCP-based 
detectors for very faint spectroscopic applications in the NUV.  
At visible wavelengths, current CCDs can achieve QEs exceeding 90\%, though
improving the detector noise characteristics will further increase the
sensitivity to the very faintest targets. The technological emphasis for
the next frontier of space-based imaging is the development of very large
format CCD mosaics to obtain wide fields of view with high spatial resolution.
In the previous section on mission concepts, we have baselined 16k $\times$ 16k
arrays as a necessary size to open huge volumes of discovery space.
However, even larger mosaics allow us to achieve wide fields with finer
pixel sampling. Tied into the development of such large arrays is the
need to attain high charge-transfer efficiencies and implement shielding
against cosmic rays, especially if the orbit is outside
the Earth's magnetosphere.

{\it Alternate material photoconductive devices} --- Newer devices that 
employ materials with a larger band-gap energy, such as GaN or diamond, 
show promise for improved mid- and far-UV QE, for lower dark noise, and 
for better red rejection.  However, the read-noise considerations for CCDs 
apply equally to these alternative materials.   If implemented in a manner 
with significant read noise, they will not supplant photon-counters for 
sensitive astronomical applications.

{\it Intensified semiconductive devices} --- An alternative path is to 
use semiconductive arrays for photon counting, so that read noise is 
not an issue.  Coupling the photoelectron from an opaque photocathode 
(deposited on a smooth metal substrate) to an electron-bombarded CCD can 
yield substantially higher QE than MCP-based photon counters.   These or 
other intensified semiconductive arrays, used for photon counting, 
promise to be extremely powerful and should receive attention for 
further development.  

\subsubsection{Microchannel Plate Detectors}

Microchannel plate (MCP) devices are currently the detector of choice 
for most UV applications.  This is primarily because these low-background 
photon counters can detect fainter sources than a CCD in any 
application with limited sensitivity, especially for observations in 
which the scientific goals permit multi-pixel binning.  However, current 
QEs for MCP-based devices are typically $\sim$10-40\%, depending on 
wavelength.  These devices still have considerable potential for 
improvement, particularly in QE (new photocathode materials and new 
substrates) and in format size.  UV astrophysics will benefit greatly 
from investments in these areas (see Siegmund 1999).

{\it Photocathode materials} --- New opaque photocathodes deposited 
directly onto MCPs are currently being tested and may lead to substantial
(factor of 2-3) QE increases in the near-UV.  Long-term stability is an 
issue and remains to be demonstrated. 

{\it MCP substrate materials} ---  Another potentially exciting 
technology, currently in a very early stage, is the so-called 
``advanced technology'' MCP, based on silicon substrates rather than 
leaded-glass, and fabricated with techniques developed in the 
semiconductor industry. In addition to other potential advantages, 
these devices may offer more hospitable substrates for high-efficiency 
photocathodes, which can be extremely contamination-sensitive.  

{\it Larger formats} --- The number of effective resolution elements in 
MCP-based detectors is limited by the maximum available size of the MCPs
and by the MCP pore spacing (currently 100 mm and $\sim$5 microns,
respectively, though not simultaneously available).  Mosaicking MCPs is 
also somewhat tricky, with substantial gaps. Improvement in any or all 
of these areas will be highly beneficial. 

\subsection{Large Lightweight Precision Mirrors}

\subsubsection{Overarching Requirements}

The science goals discussed here require large apertures, in 
addition to very efficient detectors, to achieve the requisite 
throughput. The science goals also benefit greatly from a low sky 
background, which can best be obtained in a high orbit.   Mirrors 
and their supporting structures are generally the single largest 
mass contribution to astrophysics missions, thus limiting both aperture 
and orbit.  Lightweight mirrors will allow large apertures to be 
placed into high orbit at modest cost.  

Lightweight glass mirrors currently have areal densities of 
30-40 kg m$^{-2}$, and beryllium mirrors are $\sim$20 kg m$^{-2}$.
NGST has already begun investing in large lightweight optics, and is
moving towards much lighter mirrors, with a goal of $\sim$10 kg m$^{-2}$.  
NGST and several other groups are already also making progress
with deployable segmented mirrors, to accommodate current and 
expected future launch capabilities. This work will
benefit the entire astronomical community.  

However, the relatively short wavelengths of UV-visible light will 
place much more stringent constraints on large lightweight mirrors.
Current efforts for NGST are not attempting to achieve 
performance better than diffraction-limited imaging at $\sim$1000 nm.
For UV-visible use, the mirrors will need to be more accurately
figured, to have smoother surfaces and to be non-contaminating.
If deployable mirrors are used, the alignment requirements will be 
much more stringent, and if active surfaces are required they will 
need to be controlled much more accurately.

{\it Large aperture} --- Throughput is the primary driver for large 
aperture sizes, with the moderate mission requiring an effective area
of ${\rm A_{eff} > 2 \times 10^4}$ cm$^2$ and the more 
ambitious missions requiring apertures to achieve
${\rm A_{eff} > 1 \times 10^5}$ cm$^2$ --- $\sim$100 times the 
UV throughput of HST, assuming major gains in detector QE (see
above).  Even with significant gains in detector efficiencies, these 
still imply apertures equivalent to diameters $\geq$4 m.
The most advanced concept suggested here (Class II mission) 
also has angular-resolution requirements implying an aperture of
$\sim$8 m for the primary mirror. When launch capabilities and costs 
are considered, the large aperture required for the Class II mission 
almost certainly requires a system that can be deployed from a more 
compact state.  This effective aperture and resolution must be considered
in addition to the other mirror requirements.

{\it Very good figure and alignment} --- Likely mirror materials should
be assessed for stiffness and figure accuracy. Actuation, deployment, and
alignment systems should be developed to control large optical surfaces 
to requisite accuracies for UV-optical uses. 

The Class Ib and II missions require diffraction-limited performance at
500 nm or overall wavefront error of about 1/14 waves (500 nm) rms.  
A two-optic telescope with some allowance for alignment error 
implies a primary figure error of about 1/25 waves (500 nm) rms.  
Figure error for the Class II mission primary will be divided among 
several segments, which must also be co-phased to achieve this final 
figure tolerance.  Mission Ia has a much less stringent requirement 
to produce resolution of only about 300 mas (80\% encircled energy 
in a 300 mas diameter), thus permitting a looser figure specification
(by a factor of a few) as well as more toleration of scatter.

It must be possible either (1) to manufacture the mirrors to this figure
quality with sufficient stiffness to maintain the figure, or (2) to correct
the mirror(s) to this figure quality in flight.  Alignment techniques
developed for large deployable systems (NGST) will almost certainly be
inadequate for UV/visible applications and will require considerable
refinement or development of alternative methods.  Thermal effects on the 
mirror figure must also be understood and of acceptable magnitude.
PSF correction could be accomplished actively using tertiary optics
(see below), however, multiple reflections are not desirable in the UV and
would require development of deformable gratings for a 1-bounce spectrograph.

{\it Low scattering} --- Microroughness should be determined for likely
lightweight materials. If relevant, attempts should be made to improve
surface smoothness. For both spectroscopic and imaging applications,
scattering from the mirror surfaces should be minimized, both for throughput
and for control of unwanted light in the optical path.  The current HST
mirror has a microroughness of $\sim$25 \AA\ (rms);
the current state of the art for small ($< 50$ cm) glass mirrors 
is more like $\sim$3 \AA\ (rms). Whatever technology is used to manufacture
the mirrors for the missions discussed above should be able to achieve a
microroughness better than $\sim 10$ \AA\ (rms).

{\it UV-friendly materials} --- Research into the outgassing properties and
contamination potential of likely lightweight optic materials is essential.
Contaminants, particularly molecular but also particulate, are
deleterious to UV throughput: they decrease reflectance and increase scatter.
Vigilant attention is required to assure that no part of the system, in
particular the primary mirror / assembly,  will introduce contamination onto
the optical surfaces.  Mirror substrates must also be manufactured of
materials that will accept the requisite coatings needed for UV-visible
observations.  This is particularly true if the detectors to be used require
cold windows or other contaminant attractors. 

In the following sub-sections, we mention ongoing work in this field and
comment on what additional steps will be needed for UV-optical applications.

\subsubsection{Monoliths}

Missions Ia and Ib can probably each be achieved with a single mirror.
Mission II may be built of multiple monoliths or may have a core,
high-performance monolith with smaller ``outrigger'' mirrors. 

{\it Lightweight glass or metal mirrors} --- Glass has a long track record for
precision optical surfaces and for accepting a wide variety of coatings.
Since solid glass mirrors of any significant aperture are much too massive for
spaceflight use, many creative approaches have been developed to create
lightweight glass mirrors. Lightweighting is achieved by carving out a
significant fraction of the mirror back or by casting mirrors with honeycomb
back structures and curved fronts.  An exciting recent development involves 
attaching a very thin ($\sim 1$ mm thick) curved glass face-sheet to a
honeycomb composite backing structure.  A 2-m prototype is being fabricated
under the NGST program, with apertures up to $\sim$ 4 m being considered
(Burge \& Angel 1999).  Several other types of thin mirrors are being
explored by NGST and other projects.  These are flexible, highly actuated,
and are made of materials such as Beryllium, Nickel, or glass. 

Many of the open development issues for these large thin-face mirrors will be 
addressed by the already-productive NGST and related technology studies,
including wavefront correction, actuator development, and connection of the
glass to the composite backing.  Wavefront correction will be much more
challenging for UV wavelengths than for the currently targeted 1000 nm.
However, this appears to be a very promising approach for precision UV
monoliths.

{\it Composite mirrors} ---  The possibility of manufacturing large mirrors
or mirror segments from very lightweight composites is very exciting, as it
could reach areal densities of a few kg~m$^{-2}$ (depending on mirror size),
with corresponding reductions in mass for supporting structures. Fabricating
the mirror(s) and support structure of the same materials would eliminate
themal stresses in telescopes.   Several composite mirror technologies
feature replication instead of grinding and polishing; these techniques could 
be great cost savers as well as time savers.  Double replication techniques
are being explored that would allow replication of existing (already
fabricated) telescope mirrors.

Some of the major concerns about composite mirrors are being addressed in other
development programs.  These include material behavior in a high-radiation
environment, print-through in manufacture, size of piece that can be made,
figure accuracy, mounting, thermal stability, and wavefront control. Other
major concerns are unique to the UV and require some investment before these
mirrors are validated as useful for spaceflight use.  These include the
potential for outgassing and potential self-contamination, adhesion of 
coatings, considerably better wavefront control (than currently targeted),
and surface microroughness.
   
{\it High-density cast silicon carbide} --- Although several different types
of SiC are useful for fabrication of visible wavelength mirrors, none of them
have demonstrated suitability for UV applications, having low reflectivity
and/or poor surface qualities. However, high-density cast SiC has recently
been produced that shows promise for UV mirror fabrication.  The high
reflectivity, excellent figuring and polishing capabilities, and 
low-cost manufacturing process combine to make this SiC a promising UV mirror 
material, deserving of further exploration and possible development. 
 
\subsubsection{Deployable Systems}

The more moderate science goals described above can probably be achieved with
a monolithic $\sim$4m passive mirror (perhaps with dimensions as large as
4m $\times$ 8m). However, the more ambitious science goals of the Class II
mission will require deployment of multiple segments.  Numerous concepts
exist (table folds, fixed center with petals, arrays of hexagonal mirrors,
etc.) and several prototypes are under development and testing for NGST or
other programs. We anticipate that robust, dependable, lightweight deployment
systems will be developed in these other programs and will be adaptable to use
for a large UV-optical telescope.  If these deployment schemes are to be used
for UV-visible observations, they must address usual issues of optical figure
and contamination control.        

\subsubsection{Actuators and Active Surfaces}

Actuators and active surfaces are required by future large space missions
(e.g., NGST) if they are to achieve their challenging performance goals.
Active optics are required for wavefront correction of large, lightweight
mirrors.   Actuators are also required for alignment and co-phasing between
sections of segmented mirrors. 
 
{\it Wavefront control} --- NGST will probably incorporate a flat deformable
mirror (DM) system to produce substantial figure correction of the telescope,
thereby permitting the use of a lower tolerance (and lower cost) telescope.
However, UV designs are driven to use the fewest optics possible (to avoid
throughput losses at each reflection). The extra reflections required to
produce a pupil image on the DM, and thus to provide correction over a
substantial field, lead to unacceptable throughput losses for UV applications.  
UV-optical telescope designs likely will not include an extra fold
flat for this purpose.

If the monolithic $\sim$4m primary mirrors required by Missions Ia or Ib are
lightweighted (e.g., thin face-sheets or composite shells), they almost
certainly will require active control. These systems will probably achieve the
required high accuracy figure by using actuators attached directly to the
primary facesheet, or possibly to a thin backing structure.  Actuators may
also be required to maintain a highly corrected figure throughout changing 
thermal conditions.  Large UV-optical telescopes require development of an
actuation system for monitoring and controlling the wavefront, using the
primary mirror assembly.

{\it Cophasing} --- The segmented mirror approach required for the Class II
8 m aperture must have mechanisms to allow on-orbit co-phasing
of the segments.   This co-phasing aspect is similar to several of the NGST
telescope proposals and will be able to build on NGST experience.  However,
although the Class II telescope is comparable in size to NGST, the goal
of diffraction-limited operation near 500 nm is more stringent
than the NGST near-IR goal, and
the figure correction must be correspondingly better.
Metrology and control systems required for large UV-optical telescopes are 
considerably more challenging than what is being studied for NGST.  

\subsection{UV-Optical Components and Coatings}

It is important to use every aspect of the optical system to maximize
throughput when observing faint astronomical targets.  Efficient UV optical
systems are therefore reflective (no transmissive materials are efficient at UV
wavelengths below $\sim$200 nm) and contain as few reflections as possible
(to avoid significant throughput losses at each bounce).  Throughput can be
maximized with clever optical system designs and by judicious use of 
optical materials and coatings.  New scientific capabilities may be realized
with the development of new materials or optical components.  Because of the
extreme sensitivity of UV reflectance to contamination, careful attention must
be paid to possible contamination sources (and perhaps to new cleaning
techniques) with the introduction of these new materials.

\subsubsection{Gratings}

The scientific problems discussed above all require high throughput.
If gratings are to be used in spectroscopic applications, it is
essential that they have high efficiency and produce low scatter.
Traditional mechanically ruled gratings may achieve groove efficiencies
$> 70$\% but suffer from considerable scattered light, especially at the
shortest wavelengths.  The most promising future grating technology
involves chemically etching diffractive
structures directly onto optical substrates (rather than mechanical
fabrication).  Large formats can be manufactured now, with no physical
reasons why format size cannot be substantially further increased. New 
technologies spawned from the semiconductor and micro-optics industry have
a bright future for creating UV diffraction gratings
(see Wilkinson 1999).  Future development
work is likely to be most promising in the area of aberration control,
larger formats, improved efficiency, and higher groove density. 

{\it Holographic gratings} --- Holographic gratings are today's standard
gratings for most UV spectroscopic instruments.  This is largely due to their
scattering (typically around 10$^{-5}$ \AA$^{-1}$), and to their large sizes.
Two coherent lasers are used to expose an interference pattern on a substrate
coated with photoresist.  Rulings are then formed through a chemical etching
process, creating smooth sinusoidal facets with a (first-order) groove
efficiency of $\sim$30\%.  A blaze function is then introduced, using ion
etching to ablate material and create triangular grooves. Resulting
efficiencies are currently $\sim$65\%, but are expected to exceed 70\% as the
ion-etching technology is refined, with a coincident increase in the ruling
densities.  
The holographic fabrication process allows non-parallel rulings, useful for 
controlling aberrations in the instrument light path. In the next five years,
we can expect the aberration control afforded by holographic recording
techniques to expand significantly with new techniques for introducing
aberrated wavefronts into the interference patterns. 

{\it Direct writing} --- Although direct-writing technologies are still in an
early stage of development, they are able to produce gratings with very low
scatter, efficient groove shapes, and excellent aberration correction.
The three direct-write technologies currently in development (laser writing,
e$^-$ beam writing, and excimer laser beam ablation) all use a laser or
electron beam to expose photoresist in a controlled fashion and to create
arbitrary groove shapes. Diffraction gratings using these techniques have 
been created with 85\% groove efficiencies in the visible and with scatter kept
to around $10^{-5}$ \AA$^{-1}$.

Direct writing can also create specific groove patterns for aberration control
by controlling the laser's path over the optical substrate.  This is likely to
become the preferred method to introduce aberration control, because any
computer-generated interference pattern can be drawn onto the substrate.
Current direct laser writing technologies have been used to create format
sizes in excess of 300 mm-square, but much larger sizes should be feasible.  

Today, direct-write technologies are too immature for use in high-resolution,
UV spectrometers due to low ruling densities, 100-200 g~mm$^{-1}$. However, the
industry is moving quickly and much higher groove densities are likely to be
achievable in about five years.  UV-optical astronomy stands to benefit greatly
from efficient, large, corrective gratings.

\subsubsection{Optical Coatings}

Efficient coatings are needed to maximize throughput. High reflectivities at
visible wavelengths are achieved regularly, but the situation is not
as positive at UV wavelengths, particularly below 200 nm
(see Keski-Kuha et al.\ 1999).
Aluminum has the highest intrinsic reflectance of any known material in the
UV above 100 nm, but surface oxidation can severely degrade reflectance
below 200 nm. Protective optical overcoatings can help to protect the aluminum
and optimize component reflectivity.  Coatings, particularly multilayer
coatings, can be used to maximize throughput of desired wavelengths and to
minimize the transmission of unwanted wavelengths. Surface contamination can
severely degrade performance of UV optical components, so strict cleanliness
control is required for optimum performance.  Modest gains could be
achieved
in reflective efficiency and scattering properties in the mid-UV region.
Large gains may be made in the FUV and EUV regions and in obtaining
broad spectral response.  Significant efficiency gains may be realizable
by using multilayer coatings to ``tune'' the wavelength response, to suppress
the red leak, or to form UV beamsplitters by combining coatings.

{\it Protected aluminum} --- Protective overcoatings of magnesium fluoride or
lithium fluoride can extend the useful range of aluminum mirrors to
wavelengths as short as 115 nm and 102.5 nm, respectively. Al+MgF2 is a highly
reliable coating, and in the absence of degradation caused by the deposition of
contaminants, the reflectance is stable in space. The UV reflectance
between 115 nm and 200 nm can exceed 80\% (achieved on the STIS
and COSTAR optics). LiF overcoating is hygroscopic and exhibits reflectance
deterioration and increased scatter with age. Therefore, the exposure of
Al+LiF mirrors to humidity must be controlled carefully.  Applying this
coating yields UV reflectances of 50 -- 75\% between 102.5 nm and 200 nm.

{\it Chemically Vapor Deposited SiC} --- Polished CVD SiC is the current
coating material of choice for wavelengths below 100 nm. It exhibits a high
normal incidence reflectance of over 40\% in the spectral region above
60 nm, has good thermal and mechanical properties, and provides low scatter
surfaces.  The reflectance degrades slightly with time, but can be returned
to its original value by cleaning.  CVD SiC optical surfaces must be protected
from atomic oxygen exposure in low Earth orbits. CVD SiC deposition is a
high-temperature process and not suitable for coating conventional mirror 
components or diffraction gratings.
 
{\it SiC and boron carbide films} --- The reflectance for both SiC and boron
carbide thin film coatings (ion beam deposition) is lower than that of
CVD SiC but higher than any conventional coating.
The reflectance of both materials degrades slightly with
time due to oxidation of the surface,  but stabilizes with reflectance of
over 35\% above 90 nm and 70 nm, respectively.  Acceptable performance 
can be maintained by protecting SiC coated optical surfaces from ram direction
effects.  Ion beam deposited boron carbide thin-film coating appears to be
a more robust coating than SiC, able to withstand short-term exposure to atomic
oxygen in low Earth orbit.

{\it Broad-band multilayer coatings} --- Multilayer coatings involve various
combinations of undercoating, overcoating, and doped materials.  They can be
used as filters (e.g., for red leak rejection) or for enhanced reflectivity
in specific spectral ranges.  This field is just beginning to be explored,
with significant work in materials, manufacturing, 
performance, and long-term stability still to be done.
One particularly appealing use of multilayer coatings is as beam splitters,
which could be used to increase observing efficiency of instruments
by afactor of two or better.
Other applications for beam splitters are for tracking using red light
during UV observations without requiring a separate startracker telescope. 

{\it Diamond films} --- Diamond reflectivities in the
60 nm to 100 nm range could be higher than for CVD SiC
but have not yet been realized.  Small
prototypes have been produced with acceptable smoothness for UV-visible
applications, and progress is being made in polishing diamond.  Extensive
development work is needed to produce figured diamond mirrors with sizes
and surface smoothness useful for EUV/FUV astronomical applications.

{\it Contamination and cleaning} ---  UV coatings are notoriously sensitive
to highly absorbing molecular contaminants, which cause reflectance
degradation and increased scatter. Exposure to UV light causes the contaminant
to photopolymerize. The most commonly used technique for cleaning UV optics
is a solvent flush, with the solvent depending on the coating and the
contamination (if known).  Techniques for removing photopolymerized
contamination include mild abrasive cleaning with calcium carbonate and ion
beam cleaning.   Cleaning to remove particulates is only attempted in an
emergency, as many of the coatings are soft and scratch easily; good success
has resulted from the use of CO$_2$ snowflake spray.  New cleaning techniques
will almost certainly need to be developed as new mirror materials, coating
materials, and contamination sources are incorporated into UV-optical
instrumentation.  

\subsubsection{Other Optical Components}

\noindent
{\bf Multi-object Spectroscopy} 

In spectroscopic applications, where light from stars and galaxies is
dispersed for analysis, considerable gains in efficiency
can be obtained by observing multiple objects or positions in the focal plane
simultaneously.   Multiplexing gains of over 1000 are routinely obtained in
ground-based optical/NIR multi-object-spectroscopic (MOS) instruments.   

{\it Micromirror arrays} --- Two concepts currently under development use
micromirror arrays as programmable masks.  In the ``straight-through''
arrangement, micromirrors function as an array of trapdoors,
reflecting/blocking unwanted light but opening a ``trapdoor'' for 
the desired rays.  Although this concept is still in early development,
it shows great promise for UV applications because of the clear path for the
desired photons.  In the ``selective bounce'' concept, light is reflected
either into a spectrograph (from desired objects) or into a light baffle
(unwanted).  In principle, such arrays should be coatable for use in the
FUV and even EUV, although methods need to be verified for applying UV 
coatings and for manufacturing mirrors with the requisite surface quality.   
One advantage over application of these techniques for NGST (at IR
wavelengths) is that UV-optical observations do not require operation
at cryogenic temperatures.

{\it UV-transmitting fibers} --- Many ground-based MOS instruments utilize
visible-light-transmitting fibers to feed light from selected targets
into a spectrograph. Although most optical fibers are opaque in the UV, one
type of fused silica fiber will transmit light to wavelengths as short as
180~nm.  It would be of great value to develop transmissive fibers for use at
shorter wavelengths, to reach many of the most important diagnostic spectral
lines. Considerable development work also needs to be done to couple these
fibers into a powered optical system (current lenslet arrays also do not
transmit shortward of 180 nm), to bundle them, and to understand their
flexibility properties and their behavior under extreme thermal conditions.  

\noindent
{\bf Filters}

Much of the discussion on coatings above is directly relevant to filters.
Two types of filters may be required for the missions described:
red-blocking filters and wavelength-isolating filters.

{\it Red-blocking filters} are required with many current UV-visible detectors,
particularly with CCDs.  Most red-blocking filters in use today are Woods
filters --- very thin layers of alkali metal coatings sandwiched between
layers of glass.  These generally have very low transmission in the UV
(as well as blocking the red light), and they are prone to develop pinholes
or other instabilities.  New versions of Woods filters appear to have 
considerably better stability, but the lack of UV transmission remains a
problem. 

A new type of red-blocking filter in early development, the
``nanohole filter'', consists of a thin gold film with a closely-packed array
of holes whose diameters are approximately the longest wavelength of light to
be transmitted.  Although there is still considerable work to be done in
fabricating these filters, they are expected to have good transmittance and
excellent long-wavelength rejection.  Multiple potential avenues for 
development of red-blocking filters should be explored.  

{\it Tunable filters} for broad-band and narrow-band imaging could provide
access to numerous diagnostic lines at arbitrary redshift.
At visible wavelengths, Fabry-Perot etalons with coated glass surfaces
are often employed in ground-based instruments, though application to UV
wavelengths is dubious with this technology.
Acousto-optical-tuneable-filters (AOTFs) have the potential of providing
a bandwidth selectable across $\sim$1 octave.  A tunable RF signal is applied
across a birefringent crystal, resulting in a selectable output wavelength.
AOTFs have been built and are in use at IR wavelengths on ground-based
instruments. Transmission efficiencies are not known.  For these filters to
become a viable option at UV-visible wavelengths, considerable work must be
done to characterize birefringent UV-transmitting crystals, 
followed by extensive prototype development. 

\subsection{Summary}

The science goals set forth in the beginning of this document all require
significant gains over current or anticipated UV-visible instrumentation.
These goals cannot be achieved without several major technology developments
that will not happen elsewhere:  

\begin{itemize}
\item A sustained and healthy investment in the development of high-QE,
low-noise detectors, such as the 3-D STJ/TES devices, for use at 
UV-visible wavelengths. 
\item Continued improvements to semiconductive arrays (e.g., CCDs) and 
microchannel-plate detectors, to achieve simultaneously high QE, large
format, and low noise performance. 
\item Development of large, precision, lightweight mirror surfaces with good 
microroughness properties.
\item Design and prototype development of actuation, metrology, and control  
for thin primary mirrors.
\item Continued investment in UV-visible optical components and coatings.
\end{itemize}

A number of more modest technology developments are also required or desired
and are discussed more fully in the preceding text. 

\newpage

\section{APPENDIX 1 -- Report to HST Second-Decade Committee}
 
At the UVOWG meeting at GSFC in October, Bob Brown presented an overview
of the activities and issues being considered by the HST 2nd Decade 
Committee.  He invited input from the UVOWG, and also from each of us as 
individual astronomers.  This appendix summarizes the thoughts and
suggested projects given by the UVOWG to the {\it HST Second Decade 
Committee}, regarding the importance of HST UV/O science 
and HST Key/Legacy Projects.  In fashioning the science drivers for
ST-2010 mission, it is important to think about what HST might accomplish 
in its next 11 years, both in terms of new UV/O science and in support of 
an ST-2010 mission (along the lines of HST's support of NGST). \\ 

\noindent
{\bf Start of letter to Bob Brown:} \\

At our recent UVOWG meeting (October 21-22, 1998) we discussed
with you the potential benefits of providing the HST Second Decade
Committee with suggestions for ``Key Projects" or ``Legacy Projects".
Our input would be in the context of connections between HST and
a future UV-Optical space mission (ST-2010) under consideration
by our working group. 

{\bf See http://casa.colorado.edu/$\sim$uvconf/} for material on the 
activities of the UVOWG and a preliminary description
of the ST-2010 mission. Based on discussions with members of the UVOWG, 
we list below some of our thoughts.

First, we all agree that the second decade of HST should be an
extraordinary era in UVO space astronomy, with powerful instruments
for doing UVO imaging (WFC-3, ACS, STIS) and spectroscopy (STIS, COS).
Your committee is in the enviable position of choosing how to optimize
the HST science with such instruments.  We observe that your Committee
is considering recommending implementation of large-scale ``Key Projects" 
(100-1000 orbits), to take advantage of the powerful capabilities of HST
and to simplify operations.  We also recognize the vast potential of
using HST to prepare for the significant gains in throughput available with
NGST (0.6 to $\sim20~\mu$m) and ST-2010 (0.1 to $1~\mu$m).  
The ``Legacy Projects" described below would be of significant size 
(500-1000 orbits or more, spread over several years, and would utilize 
the large databases of UV/O targets provided by the GALEX mission and 
the Sloan Digital Sky Survey.  \\  

\noindent
{\bf The UVOWG suggestions for these large projects are:} 

\begin{enumerate}
\item {\bf SPECTROSCOPIC SURVEY OF QSOs (Legacy Project, 750 orbits)}
 
One-orbit spectroscopic ``snapshots" of 500 QSOs discovered by GALEX 
and SDSS, using COS low-resolution (0.5 \AA) gratings, to attain 
S/N = 7-10 spectra.  All 500 QSOs would be observed with the far-UV 
grating (G140L).  Half of the brighter targets would be observed 
in the near-UV (G230L) from 1700 -- 3200 \AA.  The key science would be
to measure QSO rest-frame UV and EUV fluxes for qualification of the
best targets for IGM and galactic halo science. Of great importance 
is to discover which targets have damped Ly$\alpha$ and Ly-limit systems.
Some targets could be followed up with long duration HST+COS/STIS
exposures at higher resolution; however, this survey would define a
valuable sample of targets to be pursued by ST-2010.

\item {\bf QSO ABSORPTION-LINE SURVEY (Key Project, 1700 orbits minimum)}

With COS, it should be possible to perform a modest (100-target) 
QSO absorption-line survey similar to the HST/FOS Key Project, but at 
15 times better resolution ($R = 20,000$ with S/N = 20) for the 
far-UV wavelength range 1150 -- 1700 \AA.  Selected bright targets would
also be observed at mid- and near-UV wavelengths with STIS or COS.
This survey would provide a significant study of QSO absorption 
lines (H~I and metal-line systems).  The COS GTO program will
devote about 300 orbits to begin such a study, but much more 
will be needed.  The key science would include an IGM baryon 
census, study of IGM large-scale structure and its relation to 
galaxy distributions, the abundance history of the IGM, and galaxy 
halos.  For 100 QSOs with fluxes between $F_{\lambda} = 
(1-5) \times 10^{-15}$ ergs~cm$^{-2}$ s$^{-1}$ \AA$^{-1}$,   
COS could do the two far-UV settings in about 1700 orbits.  Observing 
the mid-UV settings and attaining $S/N = 30$ would require a far greater
number of orbits, and would probably require waiting for ST-2010. 

\item {\bf UVO IMAGING SURVEY OF NEARBY GALAXIES (Legacy: 1000 orbits)} 

The two unique scientific capabilities of HST - high angular resolution
and access to the vacuum-UV - should be exploited to construct a new 
``Hubble Atlas'' of galaxies. This would be of immense value in its own
right, serving to document the UVO morphology and structure of the local
galactic population. It would be of even greater value when used as a
basis for differential comparison to images of high-redshift galaxies
taken at similar rest-frame wavelengths by NGST. Finally, it would serve
as an essential pathfinder for detailed UV spectroscopic investigations
of the stellar and interstellar components of these galaxies by ST-2010.
Moderately deep images with ACS or WFC-3 at roughly 2000, 4000, and
8000 \AA\ would be acquired of a sample large enough to populate the
multi-dimensional manifold of Hubble type, absolute magnitude, 
metallicity, and effective surface brightness (several hundred galaxies). 
In order to optimize the linear resolution of the images, the nearest 
galaxies of the appropriate types should be selected, and this would 
necessitate constructing mosaics in most cases.  This sample selection 
would make it possible to resolve the brightest portion of the stellar 
population in many cases.  The UV images would be especially valuable. 
In star-forming galaxies, this light traces the youngest stellar 
population, while in galaxies with only an old stellar population, the 
UV light comes from post-main-sequence stars whose relative numbers 
and properties provide strong tests of current models of stellar 
evolution. To date, only a rather small and heterogenous sample of 
local galaxies has been imaged in this wavelength domain.

\newpage
\item {\bf UVO IMAGING SURVEY OF GAS-DYNAMICAL SYSTEMS ($>500$ orbits)} \\ 

The recent scientific literature is replete with hydrodynamic simulations 
of shock interactions with interstellar clouds and the formation and 
propagation of astrophysical jets.  Shock waves produced by supernovae 
can heat the ISM, determine the velocity dispersion of interstellar clouds, 
and govern the scale height of the ISM in galaxies.  Jets are ubiquitous 
in astrophysics and the disk/jet connection is an important component of 
numerous objects, such as young stellar objects, CVs, symbiotic stars, 
and active galactic nuclei. The radiative shocks in the nearest star
forming regions and supernova remnants are the best laboratories for
studying high mach number flows and their effects on the ISM and
star formation in galaxies.  Physical conditions such as densities and 
temperatures can be deduced by mapping the relative flux distributions 
of diagnostic emission lines in the UV and optical in radiating filaments 
using the narrow-band imaging capabilities of WFC-3, STIS, and, to some 
extent, ACS. Perhaps most importantly, the high spatial resolution and 
wide field of view afforded by HST allows us to track the motions of these
gasdynamical flows on timescales shorter than the typical radiative
cooling times.  There are hundreds of radiative shocks and filaments in 
the nearest star forming regions (e.g., Taurus-Auriga, Perseus, Orion) 
and supernova remnants for testing hydrodynamical models of jets and blast
waves. To obtain multi-epoch observations of a sample of 20 representative 
radiative flows in several of the brightest emission lines (e.g.,
H$\alpha$, [S~II], [O~III], [O~II], Mg~II) requires $\sim200$ orbits per epoch.
Programs with WFPC2 have already begun to track several flows (which
hopefully can be continued through the life of HST), however, these 
programs are generally restricted to the brightest sources and limited 
fields of view. This Key Project would seek sources with favorable geometries 
and viewing angles for revealing the physics of supersonic interactions
of the stellar ejecta with the ISM. 

\vspace{0.5cm}
Other studies might include but not be limited to:

\item {\bf STIS SPECTROSCOPIC SURVEY OF GALAXY CORES AND BLACK HOLES}

\item {\bf UVO MONITORING OF PLANETARY ATMOSPHERES AND AURORAE IN 
    THE GAS GIANTS}

\item {\bf CHEMISTRY OF THE COLD INTERSTELLAR MEDIUM}  
 
\item {\bf  UVO SPECTROSCOPIC SURVEY OF T TAURI STARS}

\end{enumerate}

\newpage

\noindent
{\bf More detailed discussions of some of the Projects follow below:} \\  

\noindent
{\bf 1. SPECTROSCOPIC SURVEY OF QSOs (Legacy Project, 500 orbits)} \\ 

The Cosmic Origins Spectrograph (COS) will provide a significant increase 
in UV spectroscopic throughput on HST.  The magnitude limit for moderate
resolution, faint-object spectroscopy will increase from about V = 15 mag
(GHRS, STIS) to $V = 17.5$ mag (COS).  Because the QSO luminosity function
rises steeply at $V > 17$, GALEX will find thousands of potential targets
for spectroscopic studies in the fields of interstellar medium (ISM),
intergalactic medium (IGM), galactic halos, quasar absorption lines
(QALs), and cosmology.  Studies of the UV continua and emission lines 
in the quasars themselves will be of significant value.  For a few 
important scientific programs (D/H, baryon census of the IGM, 
large-scale structure in Ly$\alpha$ clouds, He~II Gunn-Peterson effect, 
and metal evolution), the GALEX targets will be a treasure trove.  

The GALEX website suggests that they will find $10^6$ new quasars in the
All-Sky imaging mode.  Of these, they will obtain $10^4$ spectra; 
approximately 1000 targets will be appropriate for He~II studies
at redshifts $2 < z < 4$.  Many targets will be in the northern
continuous viewing zone (CVZ). Sorting through these targets to
find the best sources for longer-duration spectra with COS and
ST-2010 would best be done through 1-orbit ``UV spectroscopic snapshots"
with the low-resolution gratings (G140L, G230L).  In addition to
``UV-qualifying" these targets for moderate-resolution spectroscopy,
there will be significant scientific gains from this large survey:
\begin{itemize}
   
\item Measure low-$z$ QSO UV/EUV continua and emission-line fluxes 
\item Measure the QSO contributions to metagalactic ionizing background
\item Search the QSO sightlines for rare damped and Lyman-limit absorbers
\item Study numerous sightlines through galaxy halos
\item Search for ``clear sightlines" for He II studies 
\item Find partial Lyman-limit systems for D/H and metal studies

\end{itemize}
To do the really interesting science for studies of IGM and galactic 
halos, one needs to reach QSOs at $B = 18-19$ magnitude. At $B = 18$,
quasars are sufficiently abundant (about 1 per square degree) to provide 
many targets associated with galaxy halos and to map the topology of
Ly$\alpha$ clouds.  The QSO luminosity function rises rapidly 
between $B = 18-19$.   
The current flux limit for moderate resolution UV spectroscopy with
GHRS, STIS, and FUSE is $F_{\lambda} = 10^{-14}$ 
ergs~cm$^{-2}$~~s$^{-1}$~\AA$^{-1}$ ($V \approx 15$).  From 
COS sensitivity curves, we estimate that 
$19^{\rm th}$ magnitude QSOs ($10^{-16}$ flux)
could be adequately observed with COS/G140L (0.5 \AA\ resolution) in 1 orbit,
to obtain 0.015 cts/s/resolution element or S/N = 7 (higher in CVZ).
Half of the brighter targets could then be observed in the longer
wavelength band (G230L, from 1700-3200 \AA) in 250 orbits.
A dedicated survey of 750 orbits spread over 5 years would provide a 
major legacy of HST/COS that would continue into the ST-2010 era. \\ 
  
\noindent
{\bf 2. QSO ABSORPTION-LINE SURVEY (Key Project, 1700 orbits minimum)} \\

One of the original HST Key Projects involved a spectroscopic survey
of QSO absorption lines with HST/FOS (Bahcall et al.\ 1993).
That survey observed 83 QSOs with FOS ($R = 1300$) and revealed 1129  
Ly-alpha lines, 107 C~IV systems, 41 O~VI systems, 16 Lyman-limit
systems, and 1 damped Ly$\alpha$ absorber. The resolution of these
measurements is too low to derive useful abundances, kinematics, or 
physical condition information about the absorbers.  Using a subset 
of these targets (Jannuzi et al.\ 1998) the Key Project 
team derived scientific results (Weymann et al.\ 1998)
based on 987 Ly$\alpha$ absorption lines toward 63 QSOs, for lines 
complete to about 0.24 \AA\ rest equivalent width.  With COS, it
would be desirable to perform a similar survey, at 15 times better
resolution ($R = 20,000$ with S/N = 20-30) for the full wavelength
range 1150-3200 \AA.  Even doing just the two far-UV settings (1150-1700
\AA) for lines at $z < 0.2$ would be significant improvement over FOS. 

This survey would provide a significant study of QSO absorption 
lines of H I absorbers and related metal-line systems.  Towards  
QSOs located in well-studied galaxy fields at $z < 0.2$, 
it would be possible to study the baryonic large-scale structure 
in the IGM, and its relation to galaxy distributions.  
The key metal absorption lines (C~IV 1548,1551 and O~VI 1032,1038) 
can also be measured, to study the abundance history of the IGM and 
galaxy halos, and to search for metallicity gradients connected 
with the galactic winds and tidal stripping from early star formnation.

A ``modest" survey would require observing 100 QSOs with a wide 
range of redshifts.  The project would require observing QSOs with 
typical far-UV fluxes $F_{\lambda} = (1-5) \times 10^{-15}$ ---
see Table 2.3 in the 
COS proposal. The number of QSOs as large as 100 is required to build 
up statistical information about the rare Lyman-limit and damped absorption 
systems and the Lya forest systems that contain trace metals. 
As with galaxy surveys, we need many sightlines to probe the
intricate large-scale structure predicted by IGM simulations.
The Ly$\alpha$ clouds are much more numerous than $L^*$ galaxies, 
and should thus provide excellent probes of these filaments and voids.

Such a survey, in the COS far-UV band only (1150--1700 \AA), requires two 
settings per target. In the photon-counting limit, the signal-to-noise 
ratio after $N$ orbits (3000 sec each) would be:
\begin{equation}
    (S/N) = (12.2) \left[ \frac {N} {10} \right]^{1/2}
     \left[ \frac {F_{\lambda}} {10^{-15} } \right]^{1/2} 
\end{equation} 
Attaining S/N = 20--30 will require a sizeable amount of integration.
Instead of picking a single representative number for the QSO 
fluxes, it is probably best to consider a ``logarithmic distribution" 
of 10 faint QSOs ($F_{\lambda} = 1 \times 10^{-15})$, 30 with 
$3 \times 10^{-15}$, and 60 with $5 \times 10^{-15}$, 
chosen from the GALEX and Sloan targets.  Many would be
chosen in selected fields, including those studied in depth by the
Sloan survey of galaxy distributions, those in the CVZ, and QSOs in
one of the Hubble Deep Fields. To obtain S/N = 20 at R = 20,000 for 
each far-UV setting with COS requires about 27, 9, and 5 orbits, 
respectively, for fluxes of $(1, 3, 5) \times 10^{-15}$.  The ensemble of 100
targets would thus require around 1700 orbits for two FUV settings.  
Getting the longer wavelengths would add observing time, and attaining 
S/N = 30 would double the exposures. 

Thus, the ``modest" project for S/N = 20-30 would require 1700-3400 orbits, 
which is about far more time than has gone to previous individual 
HST Key projects.  Although this would be an extremely valuable IGM 
survey, it would still not address many important science questions 
that require working on fainter QSOs or require spectra at 
significantly higher S/N.  Observations of fainter QSOs would be 
needed to go after He~II absorption, to study ``double sightlines
as probes of cloud size, and to find damped Lya systems and Lyman-limit
systems, which are quite rare at low redshift. Spectra with S/N = 30  
are about the minimum one would want to use for high-quality 
studies of abundances and physical conditions. For many of the objects,
it is likely that S/N = 50 spectra would be required to pull out 
abundances of particularly important elements.
 
This exercise illustrates well the photon-starved aspect of extragalactic
UV spectroscopy with HST, even with a high-throughput spectrograph such 
as COS.  Some of the project could be done with HST, but it would require 
a huge allocation of observing time.  We conclude that we will not be 
very far along in this subject area by the time HST ends its life in 2010.

\end{document}